\documentclass[aps,prstab,graphicx,reprint]{revtex4-1}

\usepackage{graphicx}
\usepackage{dcolumn}
\usepackage{bm}
\usepackage{color}
\usepackage{amssymb}
\usepackage{amsmath}

\draft 

\begin{document}


\title{Vacuum electrical breakdown conditioning study in a parallel plate electrode pulsed DC system} 



\author{Anders Korsb\"ack}
\email{anders.korsback@helsinki.fi}
\affiliation{Helsinki Institute of Physics, University of Helsinki, P.O. Box 43, FI-00014, Helsinki, Finland}

\author{Laura Mercad\'e Morales}
\affiliation{Faculty of Physics, University of Valencia, Carrer del Dr. Moliner, 50, 46100, Burjassot, Valencia, Spain}

\author{Iaroslava Profatilova}
\affiliation{Institute of Applied Physics, National Academy of Sciences of Ukraine, 58, Petropavlivska str., 40000, Sumy, Ukraine}

\author{Flyura Djurabekova}
\affiliation{Helsinki Institute of Physics, University of Helsinki, P.O. Box 43, FI-00014, Helsinki, Finland}

\author{Enrique Rodriguez Castro}
\affiliation{CERN, European Organization for Nuclear Research, 1211 Geneva, Switzerland}

\author{Walter Wuensch}
\affiliation{CERN, European Organization for Nuclear Research, 1211 Geneva, Switzerland}

\author{Tommy Ahlgren}
\affiliation{Department of Physics, University of Helsinki, P.O. Box 64, FI-00014, Helsinki, Finland}

\author{Sergio Calatroni}
\affiliation{CERN, European Organization for Nuclear Research, 1211 Geneva, Switzerland}


\date{\today}

\begin{abstract}

Conditioning of a metal structure in a high-voltage system is the progressive development of resistance to vacuum arcing over the operational life of the system. This is, for instance, seen during the initial operation of radio frequency (rf) cavities in particle accelerators. It is a relevant topic for any technology where breakdown limits performance, and where conditioning continues for a significant duration of system runtime. Projected future linear accelerators require structures with accelerating gradients of up to 100 MV/m. Currently, this performance level is only achievable after a multi-month conditioning period. In this work, a pulsed DC system applying voltage pulses over parallel disk electrodes was used to study the conditioning process, with the objective of obtaining insight into its underlying mechanics, and ultimately, to find ways to shorten the conditioning process. Two kinds of copper electrodes were tested: As-prepared machine-turned electrodes ("hard" copper), and electrodes that additionally had been subjected to high temperature treatments ("soft" copper). The conditioning behaviour of the soft electrodes was found to be similar to that of comparably treated accelerating structures, indicating a similar conditioning process. The hard electrodes reached the same ultimate performance as the soft electrodes much faster, with a difference of more than an order of magnitude in the number of applied voltage pulses. Two distinctly different distributions of breakdown locations were observed on the two types of electrodes. Considered together, our results support the crystal structure dislocation theory of breakdown, and suggest that the conditioning of copper in high field systems such as rf accelerating structures is dominated by material hardening.

\end{abstract}


\maketitle 

\section{Introduction}
\label{section_introduction}

Vacuum electrical breakdown, also known as vacuum arcing, is the appearance of an electrically conducting plasma arc across an inter-electrode vacuum gap, following the application of a high voltage. The precise micromechanics of the phenomenon are a topic of ongoing study. Within the collaboration under which the work presented in this paper was done, the overall view of the phenomenon could be described as follows: The electric field causes an initial seed population of atoms to be emitted from the cathode, for example through the evaporation of a protrusion that locally enhances the field and causes a high local field emission current density. This seed population is then ionized by field emission electron current, accelerated back to the cathode by the field, sputtering more atoms into the gap leading to a runaway process that eventually, unless interrupted, results in a conductive plasma bridging the gap. For a more thorough description of this view along with supporting discussion, we provide references \cite{kovermann2011comparative, timko2011modelling, shipman2015experimental}. Susceptibility of a structure to have a breakdown is found to depend on the conditions of the cathode material. Threshold breakdown field strength varies greatly between different cathode metals and alloys \cite{descoeudres2009dcbreakdown}. The only material property that has been found to clearly correlate with breakdown field strength is material crystal structure \cite{descoeudres2009cobalt}, with hexagonal close packed materials being the most resistant to breakdown, followed by body-centered cubic, and face-centered cubic being the least resistant of the three.

Vacuum electrical breakdowns have in recent years become a subject of interest for the particle accelerator community due to the importance of TeV linear colliders as one of the possible directions for future high-energy physics facilities \cite{aicheler2012cdr, behnke2013ilctdr, europeanstrategy2013update, krammer2013update, aleksan2013future, nakada2015european}. For a linear collider, final particle collision energy is the product of the length and the applied accelerating gradient. Hence, there is a clear economic interest in enabling the use of the highest possible accelerating gradient. Furthermore, in other linear accelerator technologies such as medical accelerators and free electron lasers, a small machine size is an advantage in itself. Reducing the length of the machine becomes possible if accelerating gradient is increased. However, higher accelerating gradients exert stronger electric fields on the surfaces of the accelerating cavities, increasing the likelihood of a breakdown, which disrupts the accelerated particle beam \cite{dolgashev2004simulation, palaia2013beam}.

This work was carried out as part of Compact Linear Collider (CLIC), a projected future linear collider and the associated research and development project. For the reasons described above, the CLIC design specifications simultaneously require an 
accelerating gradient of at least 100 MV/m and a breakdown rate of at most $3 \times 10^{-7}$ per pulse of rf power and per metre of accelerating structure \cite{aicheler2012cdr}. On current CLIC prototype accelerating structures, this gradient requirement corresponds to a peak surface electric field of over 200 MV/m.

A newly manufactured accelerating structure requires conditioning during the first part of its operational life, starting off as highly prone to breakdowns even at low field levels and becoming capable of supporting higher fields over time \cite{brown1989status, adolphsen2001processing, adolphsen2003normal, rodriguez2007_30ghz, catalanlasheras2014experience, degiovanni2016conditioning}. For this reason, controlled conditioning of structures is part of the construction and commissioning of a linear accelerator. Input power is increased gradually in response to breakdown behaviour until a desired performance target has been met. In recent years, a number of prototype high-gradient accelerating structures have been subjected to long test runs in accelerator labs around the world \cite{matsumoto2011high, degiovanni2014high, catalanlasheras2014experience, degiovanni2016conditioning, zennaro2017high, wuensch2017high}. The structures were run for thousands of hours of runtime and subjected to hundreds of millions of rf pulses, had thousands of breakdowns in the process, and saw large improvements in breakdown performance. For example, the prototype CLIC accelerating structure TD26CC \cite{grudiev2010design} was tested at CERN in 2014 \cite{degiovanni2014high}. Over the course of 1800 hours of runtime, its breakdown rate decreased from $7 \times 10^{-5}$ to $2 \times 10^{-5}$ per pulse, even as its accelerating gradient was increased from about 30 MV/m to 105 MV/m and its pulse length was increased from 50 ns to 250 ns. 

The ability to improve structure performance to such an extent by conditioning, and the long runtimes required to do so, make conditioning a subject of interest for the accelerator community. If each accelerating structure needs to be individually conditioned for months or more as part of its production process, there is an economic interest in speeding up the process and reducing the time needed. 

Understanding the physics of conditioning could show ways to improve conditioning rates and reduce the runtime needed to reach conditioning goals, or even reach higher levels of ultimate performance. If the conditioning rate is found to depend significantly on material properties (which we will show to be the case in this paper), this would imply that choice of materials and treatments are important considerations in accelerating structure design. If conditioning rate depends significantly on how rf power is applied to a structure, that would imply that efforts should be made to optimize conditioning algorithms. Furthermore, an understanding of the conditioning process may lead to an improved understanding of the breakdown process itself.

Beyond the accelerator field, conditioning could be a relevant topic for any technology where breakdown causes device failure, either by directly disrupting device operation or by causing cumulative hardware damage. For instance, if a particular device makes use of vacuum for electric insulation, controlled conditioning could reduce needed vacuum gap sizes, enabling device miniaturization.

Conditioning of an accelerating structure can either be quantified as a decrease in breakdown rate for a given accelerating gradient and rf pulse length, or as an increase in the accelerating gradient and/or rf pulse length that result in a given breakdown rate. To be able to quantify conditioning progress over the course of an experiment where more than one of these parameters change over time, and to be able to compare conditioning progress between different experiments, the concepts of \emph{normalized breakdown rate} and \emph{normalized gradient} have been established \cite{degiovanni2016conditioning}. It has been found that the breakdown rate per pulse is proportional to the accelerating gradient $E$ and pulse length $\tau$ according to the power laws\cite{grudiev2009new}: 

\begin{center}
\begin{equation}\label{equation_powerlaw_E_tau}
$$
$ \text{BDR} \varpropto E^{30}\tau^5 $.
$$
\end{equation}
\end{center}

Consequently, one can define a normalized breakdown rate as

\begin{center}
\begin{equation}\label{equation_normalized_bdr}
$$ 
$ \text{BDR}^* = \dfrac{ \text{BDR} }{  E^{30}\tau^5 } $
$$
\end{equation}
\end{center}

and a normalized accelerating gradient as

\begin{center}
\begin{equation}\label{equation_normalized_gradient}
$$
$ E^* = \dfrac{ E \tau^{1/6} }{  \text{BDR}^{1/30} }  $.
$$
\end{equation}
\end{center}

These normalized breakdown performance quantities are equivalent descriptions that serve the purpose of providing a metric of comparable breakdown performance, or "conditioning state", in a way that adjusts for differences in breakdown rates, applied accelerating gradients and pulse lengths. Such comparison has recently been performed for the conditioning experiments carried out at different accelerator institutes \cite{degiovanni2016conditioning}. This has led to the conclusion that conditioning is mostly or entirely caused by the rf pulses themselves, regardless of whether or not there is a breakdown. This conclusion implies that conditioning is caused by physical changes within the structure that occur when it is exposed to strong electric fields.

At CERN, prototype CLIC accelerating structures are tested at the klystron-based testing facilities Xbox-1 to Xbox-3 \cite{jacewicz2011instrumentation, catalanlasheras2014experience, woolley2015high, woolley2015high_phd}. In addition, CLIC employs a complementary fundamental breakdown research effort using the DC Spark Systems \cite{kildemo2004new, kovermann2011comparative, rajamaki2014breakdown, shipman2015experimental}. These are experimental setups that apply DC voltages in the kV range over electrode pairs separated by a vacuum gap in the size range of tens of micrometers. Fast switching allows the voltage to be pulsed in the microsecond pulse length range at a repetition rate up to 1 kHz \cite{soares2012pulsegenerator}. This allows the electrodes to be subjected to electric field strengths and pulse lengths of the same or nearby orders of magnitude as those that rf accelerating structures are subjected to during operation. Such systems thus allow fundamental experimental studies to be carried out at low cost and high experimental throughput compared to the Xbox facilities.

In this paper, we present the results of conditioning experiments we performed in the Large Electrode DC Spark System \cite{rajamaki2014breakdown, shipman2015experimental}. It is a DC spark system that applies voltage pulses over a pair of parallel disk electrodes of 62 mm diameter, subjecting their surfaces to a uniform electric field. This system has shown itself to be analogous to an rf accelerating structure with regards to breakdown rate dependence on field strength and pulse length \cite{shipman2015experimental}, and with regards to statistics of breakdown occurrence \cite{wuensch2017statistics}. The large cathode surface subjected to electric field makes this system particularly well suited for conditioning experiments. Similarly to an rf structure, the total surface subjected to field is orders of magnitude larger than a breakdown spot, i.e. the area directly damaged by the breakdown. Hence, during a conditioning run, a large fraction of the surface never experiences a breakdown, but is nevertheless affected by the pulsing.

For our conditioning experiments, we used copper electrodes that have been subjected to the same thermal treatments as the prototype CLIC accelerating structures are as a part of the manufacturing process currently in use at CLIC \cite{wang2010fabrication}. We did this in order to obtain results that are as comparable as possible with the results of CLIC rf conditioning experiments. These treatments heat the copper structures close to the melting point of copper of 1040 $^{\circ}$C, resulting in a large grain size (up to a few mm in diameter or more) polycrystalline copper of low hardness, referred in the remainder of the article as "soft electrodes". In addition, we did comparative experiments on as-prepared, diamond-turned electrodes of "harder copper. These electrodes were not subjected to any thermal treatment, thus the residual stresses after the machine-processing are not relaxed. These electrodes are referred in the remainder of the article as "hard electrodes." Both types of electrodes were tested in order to study the effect of the material's intrinsic mechanical properties on the conditioning process.

\section{Experimental methods}
\subsection{Sample preparation}
\label{subsection_sample_preparation}

\begin{figure}
\includegraphics{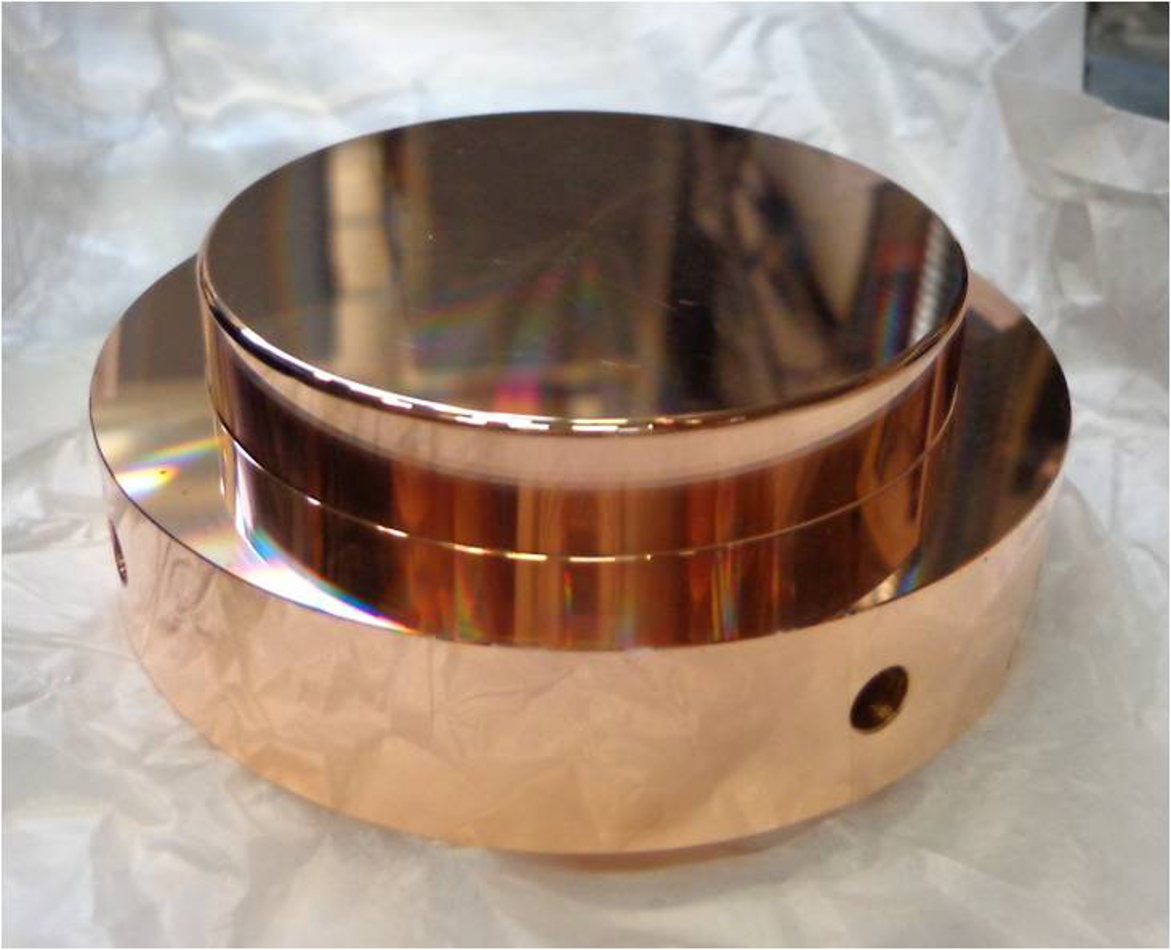}
\caption{An electrode used by the Large Electrode DC Spark System. The inter-electrode vacuum gap is formed between the top surface of the electrode, and the same surface on an identical top electrode facing it symmetrically. The outer rim of the electrode supports the ceramic spacer that holds the electrodes apart and determines the vacuum gap distance. Electrode edges are gently rounded (rounding radius $\gg$ gap distance) to prevent breakdown through the ceramic, and to reduce field enhancement by the edges. }
\label{Fig:largeelectrode_photo}
\end{figure}

The sample used for each conditioning experiment was a pair of interchangeable, identically shaped and treated copper disk electrodes. The electrodes have an outer rim running around the cylinder side, see Fig. \ref{Fig:largeelectrode_photo}. A parallel electrode surface geometry is secured by stacking the electrodes facing each other, separated by a cylindrical ceramic spacer resting on the rim of the bottom electrode and, in turn, supporting the top electrode. When thus stacked, the electrodes face each other with aligned parallel disk surfaces with a diameter of 62 mm. The size of the inter-electrode vacuum gap is determined by the height of the spacer, and was 60 $\mu$m in these experiments.

The material used for the samples is 3-D forged oxygen-free electronic (OFE) copper, in accordance with the CERN specification for copper used for accelerating structures \cite{cern_accelerating_structure_copper}, which corresponds approximately to UNS C10100 Grade 1 of the ASTM B224 standard classification of coppers. Key parameters are a maximum oxygen content of 5 ppm, a maximum grain size of 90 $\mu$m and a minimum Brinell hardness number of 60. The main steps of the production process for accelerating structures are \cite{wang2010fabrication}:

\begin{enumerate}

\item	Final machining of the parts of the structure is made with single crystal diamond tools on micrometer precision lathes and mills.
\item	The parts are bonded together into structures with a heat treatment at temperature up to 1040 $^{\circ}$C in a 1 atm hydrogen atmosphere. This joins together the parts without leaving a seam, welding joint or other interface between them. This process takes about 7 hours, during which the top temperature is maintained for 3-4 hours.
\item The bonded structures are subjected to a vacuum bake-out at 650 $^{\circ}$C lasting 4-5 days in order to remove hydrogen introduced during the bonding.

\end{enumerate}

The electrodes were prepared in a way matching this production process as closely as possible. As per the first step, all electrodes were machined identically, as a single piece each, with all critical surfaces and dimensions manufactured to micrometer tolerance. Some electrodes were left in that machined state, these are the ones we refer to as the "hard electrodes". The others were put through the two subsequent steps, which change their material state significantly, and became the electrodes we refer to as "soft electrodes". As each soft electrode was already machined into its final shape in the first step, no actual bonding of copper parts to each other took place during the second step. Rather, each electrode was put individually through the process in order to make it undergo the same change in material state as accelerating structure parts do as part of this step.

\subsection{Experimental setup and procedure}

A description of the Large Electrode DC Spark System is provided here. It consists of a chamber housing the electrode setup in ultra-high vacuum, with electrical feedthroughs for voltage application. It is connected to a pulse generator system \cite{soares2012pulsegenerator} that applies voltage pulses over the electrodes at a high repetition rate, and which in turn is controlled by higher-level control software on a desktop computer.

The pulse generator uses a DC power supply to charge a coaxial cable that acts as a pulse forming line (PFL) and as a capacitive charge buffer. Voltage pulses are applied from the PFL over the electrodes by closing and opening a fast, high-voltage solid state switch manufactured by Behlke. The system is capable of pulsing at a rate of up to 1 kHz, subject to power dissipation constraints at higher voltages. When a breakdown happens between the electrodes, the PFL is drained of charge, producing an approximately constant breakdown current during 2 $\mu$s, after which it is depleted and the breakdown plasma dies out. The system detects a breakdown by measuring the current going to the electrodes, and uses the exceeding of a threshold value to determine that a breakdown took place. The switch is controlled by a digital signal provided by an embedded microcontroller, which counts the number of pulses applied and stops pulsing when a breakdown is detected. 

Higher-level experiment control takes place in LabVIEW on a lab PC communicating with the microcontroller. Each conditioning experiment is started by manually adjusting the pulse voltage to find the lowest voltage at which breakdowns occur. When a suitable voltage to start at has thus been found, the main experimental control algorithm is started, an algorithm which mimicks as closely as possible the algorithm currently in use at the Xbox facilities through which input power is controlled in conditioning experiments \cite{catalanlasheras2014experience}. Each iteration of the algorithm, pulsing is started by ramping up voltage asymptotically towards a set value, so that 90\% of the set value has been reached after about 700 ms and 99\% after about 1400 ms. Pulsing continues until either a breakdown happens or a certain max number of pulses $n_{max}$ have been applied during this iteration. Then, pulsing is paused and the voltage set value is changed for the next iteration as follows:

\begin{itemize}
\item If no breakdown happened, voltage is increased by an amount $\Delta_+$
\item If a breakdown happened after more than a minimum number of pulses $n_{min}$, voltage is kept unchanged
\item If a breakdown happened after $n$ pulses and $n < n_{min}$, voltage is reduced by ${(1-n/n_{min}) \Delta_-}$

\end{itemize}

In all conditioning experiments, the following parameter set was used: $n_{min} = 20000, n_{max} = 100000, \Delta_+ = \Delta_- = 10$ V. In all experiments, a constant voltage pulse length of 16.7 $\mu$s was used.

For a more detailed description of the system, including a schematic of the circuit and a discussion of instrumentation issues, we provide references \cite{kovermann2011comparative, soares2012pulsegenerator, rajamaki2014breakdown, shipman2015experimental, wuensch2017statistics}. 

\subsection{Sample post-mortem microscopy}

It was thought that the spatial distribution of breakdown spots on the surface of the cathode could be relevant for drawing conclusions about the effect of the conditioning process on the sample. To obtain such information, samples 4 and 5 were subjected to a post-mortem surface analysis using an optical microscope (Zeiss Axio Imager). This microscope is equipped with a motorized scanning stage with a spatial movement resolution of 0.1 $\mu$m and a reproducibility of  $\pm$ 1 $\mu$m. It is connected to a desktop computer equipped with the latest Zeiss image analysis software (ZenCore2). The microscope scans the entire disk surface and stitches together a reconstructed global image from individual images. The stitching is done with an overlap of 10\% to guarantee the fidelity of the reconstructed image. The fine machining tolerance of the sample surface causes it to be very smooth and highly reflective, except for where it has been disturbed by breakdown. Thus, when the sample surface is illuminated, a sharp contrast appears between the breakdown sites and the surrounding undisturbed surface. This allows the analysis software to identify and count breakdown sites, and to calculate the ratio of area modified by breakdown to total sample area.

\section{Results}
\subsection{Conditioning progress}
\label{subsection_conditioning_progress}

Five conditioning experiment runs were carried out, each on a fresh, previously unused pair of electrodes. We refer to these pairs as Sample 1 through Sample 5, in the chronological order that the measurements were carried out. Samples 1, 3 and 5 were soft electrodes subjected to the thermal treatments described in section \ref{subsection_sample_preparation}. Samples 2 and 4 were untreated hard electrodes. Samples 1, 2 and 3 were all from the same batch of machined electrodes. Samples 4 and 5 were both from the next batch. Thus, it is reasonable to assume that the soft and hard electrodes were of near-identical material state before the softening thermal treatments were carried out.

\begin{figure}
\includegraphics{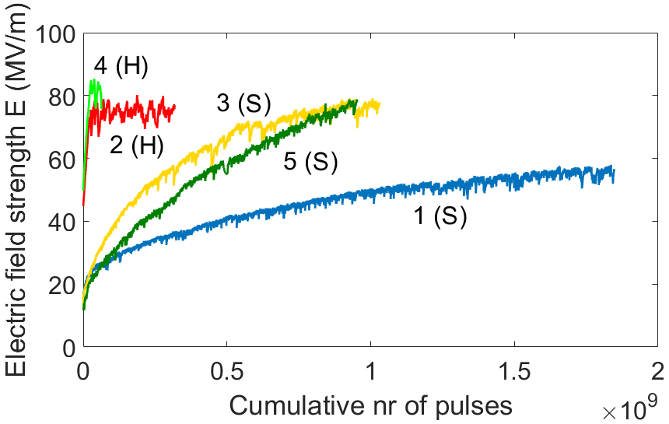}
\caption{Electrode surface electric field strength over the course of all five measurement runs, determined by dividing the applied voltage with the gap distance of 60 $\mu$m. Series are labeled in figure by sample number and type, (H)ard and (S)oft.}
\label{Fig:E_vs_cumnrpulses}
\end{figure}

Figure \ref{Fig:E_vs_cumnrpulses} shows the progress of the experiments, in terms of electric field strength as a function of cumulative number of pulses applied to each sample. Each measurement run was continued for as long as was considered relevant, considering competing hardware requirements of other experiments. An exception is the measurement of Sample 5, which was unfortunately interrupted by hardware failure. Electric field strength is simply determined as applied voltage divided by the gap distance 60 $\mu$m, i.e. actual local field strength at a breakdown site could be greater due to geometric field enhancement from irregularities. In an rf accelerating structure, surface electric field strength is proportional to accelerating gradient, thus we will use electric field strength as analogous to accelerating gradient in our analysis. As pulse length was constant during all measurements and the breakdown rate was effectively kept around 1.5 $\times 10^{-5}$ in all experiments, electric field strength is practically proportional to normalized field strength (Eq. \ref{equation_normalized_gradient}), thus we do not separately consider the latter.

As can be seen, the two hard samples conditioned relatively quickly to an ultimate value of field strength, after which electric field strength stably fluctuated around the ultimate value but with no further conditioning taking place. The saturated field observed for Sample 4 was slightly higher than that of Sample 2. Visual inspection of Sample 2 after its measurement run had ended showed that almost all breakdowns had happened near the edge of the electrode, clustered on a single spot. Later analysis of the system provided an explanation for this clustering and for the stabilization of electric field strength at a lower level than for Sample 4. The rounding of the electrode edge, despite having a radius orders of magnitude larger than the inter-electrode gap, creates field enhancement at the point on the surface where the rounding starts, due to a discontinuity in the second derivative of the contour of the surface. This effect is particularly prominent when the electrodes are misaligned by being parallel but off-centre relative to each other. Then, a part of the cathode edge faces a flat anode surface, rather than a matching edge, further increasing field enhancement. Thus it is likely that there was a discrepancy between actual and nominal field strength in the area where most breakdowns of Sample 2 happened, and that the actual field strengths in the areas where the breakdowns happened on the respective samples are closer to each other than the nominal field strengths shown in Figure \ref{Fig:E_vs_cumnrpulses}.

The soft electrodes, on the other hand, showed much slower, steadier conditioning. The surface of Sample 1 was unfortunately contaminated by aluminium oxide particles during the heat treatment, but used nevertheless, allowing for our experiment to also test the effect of surface contamination on conditioning. Sample 1 conditioned very slowly. Samples 3 and 5, for which the heat treatment was successful, conditioned much faster. As can be seen in Figure \ref{Fig:E_vs_cumnrpulses}, field strength in these conditioning experiments behaved very similarly against the cumulative number of pulses. However, Sample 3 conditioned slightly faster, saturating at roughly the same field strength as the hard Samples 2 and 4 did. Subsequent measurements using Sample 3 (not included in this paper) showed no signs of further conditioning during another 1.7 billion pulses. Sample 5 conditioned slightly slower. This could either be due to small differences in material condition between the two, or due to a scratch that appeared on both electrodes of Sample 5 during system assembly. It is interesting to note, however, that Sample 5 shows no signs of saturation even at the end of its run, when the field strength has reached the same level as the saturation level in Sample 3.

A better understanding of the conditioning of the soft samples can be gained by considering the normalized breakdown rate (Eq. \ref{equation_normalized_bdr}). Comparisons between different rf conditioning experiments have shown that conditioning progress tends to follow a power law\citep{degiovanni2016conditioning}:

\begin{center}
\begin{equation}\label{equation_powerlaw_nBDR}
$$
$ \text{BDR}^* \varpropto n_p^A $,
$$
\end{equation}
\end{center}

\begin{figure}
\includegraphics{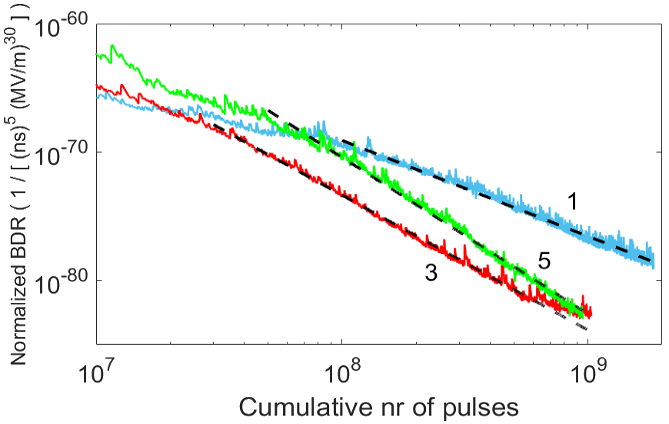}
\caption{Normalized breakdown rate (Eq. \ref{equation_normalized_bdr}) over the course of the respective measurement runs of Samples 1, 3 and 5 (labeled in figure), along with fits for conditioning gradient. For Sample 3, fit is plotted beyond fit range to show divergence from power law due to saturation. Unit of normalized BDR chosen consistently with other work \citep{degiovanni2016conditioning}.}
\label{Fig:nBDR_vs_cumnrpulses}
\end{figure}

\begin{table}
\begin{center}
\scalebox{0.9}{
  \begin{tabular}{| c | c |} 
    \hline
    \textbf{Structure} & \textbf{A} \\
     \hline
    Sample 1 & -7.4 \\
    Sample 3 & -10.5 \\
    Sample 5 & -12.2 \\
    \hline
    TD26CC & -9.2 \\
    TD24R05{\#}2 & -6.8 \\
    TD24R05{\#}4 & -8.0 \\
    \hline

  \end{tabular}
  }
\end{center}
\caption{Power law exponents for normalized breakdown rate as a function of cumulative number of pulses, in our DC conditioning experiment and in comparable rf conditioning experiments \cite{degiovanni2016conditioning}.}
\label{tableitem:powerlaw_exponents}
\end{table}

where BDR$^*$ is normalized breakdown rate, $n_p$ is cumulative number of pulses, and $A$ is a fitting parameter. For this reason, it is instructive to plot normalized breakdown rate as a function of cumulative number of pulses using logarithmic axes, whereby a power law dependence shows as a straight line with a gradient proportional to the exponent of the power law. Figure \ref{Fig:nBDR_vs_cumnrpulses} thus shows normalized breakdown rate for Samples 1, 3 and 5 along with fits of the power law (Eq. \ref{equation_powerlaw_nBDR}) to illustrate the power law dependence. Table \ref{tableitem:powerlaw_exponents} shows the power law exponents obtained by the fit, along with the exponents obtained in the aforementioned rf conditioning study \cite{degiovanni2016conditioning} for comparison. We see that the power law fits for all the soft samples over more than the last order of magnitude of number of pulses. We further see that the normalized breakdown rate of Sample 3 starts deviating from the fit at around 600 million pulses, visibly levelling off in a way that suggests saturation. Samples 1 and 5, on the other hand, do not show any signs of saturation yet.  The exponents obtained in our experiments are in our opinion remarkably close, if somewhat higher, than those obtained from the rf conditioning experiments.

\subsection{Spatial distribution of breakdowns}
\label{subsection_spatial_distribution_of_breakdowns}

\begin{figure*}
\includegraphics[width=1.0 \textwidth]{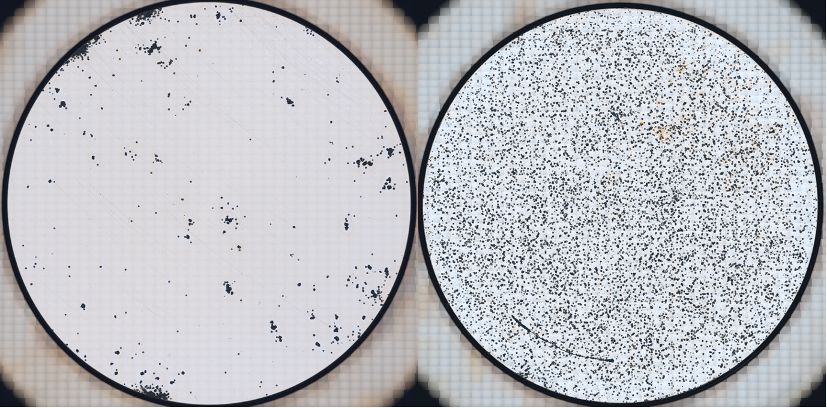}
\caption{Micrographs obtained by a scanning optical microscope of a hard electrode (Sample 4, left image) and a soft electrode (Sample 5, right image). Illumination of the samples provides a sharp contrast between the smooth undisturbed surface (white) and the areas damaged by breakdown (black).}
\label{Fig:sem_samples_45}
\end{figure*}

Figure \ref{Fig:sem_samples_45} shows the spatial distribution of breakdown spots on the cathodes of Sample 4 and 5. The correspondence between identified spots and breakdown is imperfect since a single breakdown could plausibly cause several non-contiguous areas to be damaged, or a breakdown could happen on an already-damaged area rather than create a new one. However, the difference between the micrographs of the hard Sample 4 and the soft Sample 5 is clear enough that the imperfect correspondence should not affect the interpretation of the results.

Sample 5 has a very even distribution of breakdown spots, with no visible clustering or preponderance anywhere. The only exception to this is the scratch, which seems to have a contiguous string of breakdown spots on the left end. In section \ref{subsection_conditioning_progress}, we attributed the slower initial conditioning of Sample 5 relative to Sample 3 to the scratch. The micrograph is thus consistent with this interpretation, even if we cannot know during what part of the experiment the breakdowns on the scratch occurred.

On Sample 4, on the other hand, breakdowns are strongly clustered, to the extent of forming several contiguous areas of connected spots, and large areas almost free of breakdown spots. The largest clusters on Sample 4 are visibly larger than any clusters on Sample 5, which is significant when considering that Sample 5 has had a larger total number of breakdowns on it by a factor 12.3. In the field of breakdown study, it is known that breakdowns have a general tendency to be temporally clustered \cite{shioiri2000dielectric, adolphsen2003normal, shipman2015experimental, wuensch2017statistics}. In previous work, we showed that such temporal clustering correlates with spatial clustering \cite{wuensch2017statistics}, and identified two kinds of breakdowns: Primary breakdowns that occur independently, and follow-up breakdowns that are induced by preceding breakdowns (and thus are the second and subsequent breakdowns in a temporal cluster). However, a quick look at the measurement data for Sample 4 showed that 54\% of all breakdowns were not part of temporal clusters, and only 2.3\% of all breakdowns were part of temporal clusters with six or more breakdowns (for this purpose, two subsequent breakdowns were considered part of the same temporal cluster if there were less than 5000 pulses between them). Thus, a simple look at the micrograph of Sample 4 shows that the spatial clustering is much stronger than the temporal.

Furthermore, the three largest clusters on Sample 4 are all on the edge, likely a consequence of the edge field enhancement described in Section \ref{subsection_conditioning_progress}. Furthermore, two of these three large edge clusters are close to each other, but the third is almost opposite to the other two. Sample 5, however, shows no recognizable edge clusters.

We will further consider the implications of these distributions in combination with the conditioning progress of the samples in the next section.

\subsection{Discussion}

Of the samples on which we conducted measurements, three saturated at a level of ultimate performance, and did so at levels close to each other, all of them at field strengths around 75-80 MV/m. These were the two hard Samples 2 and 4, and the soft Sample 3. The hard samples reached that level after about 100 breakdowns over 25 million pulses each. After this, no further conditioning of these samples was observed. Sample 3 started showing signs of saturation only after about 600 million pulses, and stabilized at the level of ultimate performance after about 900 million pulses. Post-mortem microscopic analysis (Figure \ref{Fig:sem_samples_45}) showed that the spatial distribution of breakdowns is highly clustered on the surface of the hard Sample 4, while evenly distributed over the entire surface of the soft Sample 5.

These observed differences in the behaviour of hard and soft electrodes support a few hypotheses about the nature of the conditioning process. One is that conditioning is mainly a process of material hardening. This would explain why all electrodes that saturated at a level of ultimate performance did so at about the same level. The soft electrode gradually assumed the same hardness, and thus breakdown resistance, as the hard electrodes. The hard electrodes, on the other hand, were largely unaffected by the pulsing, with the exception of the initial rapid rise of breakdown performance.

We also observe that the evolution of normalized breakdown rate (Eq. \ref{equation_normalized_bdr}) in the soft samples corresponds closely to that in comparable rf structures \cite{degiovanni2016conditioning}. The power law identified for rf structures (Eq. \ref{equation_powerlaw_nBDR}) holds true for all three of our soft samples, and the power law exponents we obtained are similar, if slightly higher, than those of rf conditioning experiments (Table \ref{tableitem:powerlaw_exponents}). This strongly indicates that the changes in the samples that underlie the conditioning process are the same in both cases, that our soft DC electrodes are indeed experimentally analogous to rf structures for the purpose of conditioning.

The clear difference in breakdown spot distribution between the hard Sample 4 and the soft Sample 5 (Figure \ref{Fig:sem_samples_45}) suggests that the pulsing has a qualitatively different effect on the two samples. That is, the rapid rise of the hard samples to the level of ultimate performance and the much slower rise of the soft samples are not the same process happening at different rates, but processes that are different in kind. For the hard samples, the rise was likely due to the removal of oxide or other surface impurities. This explanation is consistent with longstanding experience of running rf accelerating structures, and temporarily exposing them to air during machine maintenance. The exposure to air causes a sharp dip in structure breakdown performance, followed by a recovery of performance back to pre-exposure levels during subsequent operation. Our samples would, likewise, have obtained such a temporary condition of decreased breakdown performance due to exposure to air prior to their installation into the experimental setup. In the case of our soft samples, this removal of impurities would also have taken place but been obscured by the much slower conditioning of them through hardening.

For the highly clustered breakdown spot distribution of Sample 4, we propose the following explanation: Breakdowns on a hard, highly breakdown-resistant surface cause a local decrease in breakdown performance due to damage done to the surface. This local decrease in breakdown performance lasts longer than the timespan of a temporal breakdown cluster, and is possibly even permanent. That is, a primary breakdown is not only prone to induce immediate follow-up breakdowns, but also increases the probability that future primary breakdowns occur in the area. This would explain why Sample 4 shows much stronger spatial than temporal breakdown clustering, as explained in Section \ref{subsection_spatial_distribution_of_breakdowns}.

Sample 5, on the other hand, starts with a soft surface highly prone to breakdown, and conditions towards higher hardness over time. This could counteract breakdown spatial clustering in several ways. A soft surface is more susceptible to breakdown than a hard surface, so it might be that surface damage caused by breakdown has less of an effect on breakdown susceptibility on a soft than on a hard surface. It might also be that the breakdown susceptibility of an area damaged by breakdown is largely or entirely unaffected by the hardness that the area had prior to being damaged. In that case, the absolute breakdown susceptibility of a damaged area would be similar in a hard and a soft sample, but the soft sample would have a relatively higher breakdown susceptibility on undamaged areas of the surface, causing a larger fraction of all breakdowns on the soft sample to happen in new, undamaged places, and thus breakdowns to be less spatially clustered.

Furthermore, it could be that conditioning through hardening is a self-regulating process. That is, regions on the surface that have a high probability of breakdown during operation also condition faster, causing local breakdown rates at different parts of the surface to converge. This could counteract any longer-term effect that local surface damage would have, as well as counteract possible other effects that could cause a non-uniform breakdown spot distribution. This effect would not have been present in Sample 4, as it was hard from the start and practically did not condition through hardening. This explanation is thus consistent with the remarkably uniform breakdown spot distribution on Sample 5 and with the non-uniform distribution on Sample 4. It is also consistent with observation of breakdown spot distributions on CLIC accelerating structure cell irises \cite{degiovanni2016postmortem}. Electric field strength on the iris surface is highly uneven. Considering this and the steep power law dependence of breakdown rate on field strength (Eq. \ref{equation_powerlaw_E_tau}), one might expect almost all breakdowns to happen in the regions of highest field strength. While breakdown spots on the irises were far from as uniformly distributed as on Sample 5, the study found no clear correlation between field strength and breakdown spot density, suggesting that the effect of the stronger local field had at least in part been counteracted by faster local conditioning. A similar situation of uneven field strength arguably exists in our experiment due to the edge field enhancement described in Section \ref{subsection_conditioning_progress}. Sample 4 that did not condition through hardening had three large edge clusters, likely a consequence of the stronger field at the edges. Sample 5, on the other hand, showed no such edge clusters, suggesting that the effect of the stronger edge field was counteracted by conditioning in the self-regulating manner described above, similarly to the structure cell irises.

Recent theoretical and computational efforts to model and understand the breakdown phenomenon \cite{nordlund2012defect, engelberg2018stochastic} suggest an underlying mechanism for our interpretation of our results: Movement of dislocations in the crystal structure of the cathode. Such movement is caused by stress generated by the applied electric field, and is thought to play a key role in the breakdown process by causing the growth of tiny protrusions on the cathode surface, particularly through interaction with near-surface defects such as voids or precipitates \cite{pohjonen2011dislocation, pohjonen2013dislocation, muszynski2015dislocation}. Such protrusions cause local field enhancement and a high local field emission current density, causing atoms to be evaporated from the protrusions \cite{parviainen2011electronic}. This provides both an initial population of neutral atoms in the inter-electrode gap and the electron current needed to ionize them \cite{timko2011onedimensional}, starting the buildup process of breakdown plasma described at the start of Section \ref{section_introduction}.

This mechanism provides an explanation for why breakdown resistance is correlated with material hardness, why soft electrodes condition through hardening, and why conditioning through hardening is self-regulating as we have hypothesized. Differences in hardness between samples of the same metal are mainly caused by differences in dislocation mobility \cite{hull2001introduction}. Hence untreated copper is hard because it contains many obstacles to dislocation movement, such as grain boundaries and crystal structure defects \cite{hull2001introduction, hall1951deformation, petch1953cleavage, cordero2016sixdecades}. Dislocations themselves are one such obstacle, especially when locked in place by other dislocations \cite{hull2001introduction}. Mechanical stress is known to introduce dislocations into metal \cite{hull2001introduction}. Hence, when a soft copper sample is pulsed with voltage, new dislocations are created and existing dislocations can move toward and accumulate near the surface. This causes hardness to increase locally, similarly to a sample undergoing work hardening. The higher the density of dislocations, the higher the probability that a moving dislocation will be locked in place by a dislocation already present, preventing it from reaching the surface and contributing to the formation of a new field emitting protrusion. Hence, a higher dislocation density, i.e. a higher hardness, inhibits dislocation movement more strongly, causing a stronger electric field to be required to achieve the needed dislocation movement for both breakdown and for further hardening. Hence, breakdown resistance is positively correlated with hardness, and local breakdown probability and local conditioning rate positively correlated with each other by both of them being dependent on the rate of dislocation movement.

\section{Conclusions}

We have conducted pulsed DC breakdown conditioning experiments on parallel plate copper disk electrodes. We used three pairs of soft electrodes subjected to the same thermal treatments as rf accelerating structures are subjected to in the CLIC study at CERN \cite{wang2010fabrication}. We found that the conditioning progress of our electrodes is described by the same kind of power law that describes the conditioning of the comparable rf structures \cite{degiovanni2016conditioning}. We obtained similar, if slightly higher, values for the power law exponents that quantify conditioning rate.

This result suggests that the power law has some measure of generality. Furthermore, it suggests that our system is experimentally analogous to an rf accelerating structure for the purpose of conditioning, and thus provides a low-cost, high-throughput way to test the effect of different material conditions and operating algorithms on the conditioning process.

We also conducted conditioning experiments on two pairs of hard electrodes that had not been subjected to the softening material treatments. We found that these electrodes quickly reached a level of ultimate performance, something that only one soft electrode did, and the soft electrode did so much slower. This soft electrode pair and both hard electrode pairs reached approximately the same level of ultimate performance. We further investigated the spatial distribution of breakdowns on one hard and one soft cathode through post-mortem microscopy, and found breakdowns to be very evenly distributed on the surface of the soft cathode but highly clustered on the surface of the hard cathode.

These differences in the behaviour of soft and hard electrodes suggest that conditioning is a process of microstructural hardening, that the soft electrode pair reached its level of ultimate performance by converging towards the hardness of the hard electrode pairs. This result is significant for a number of reasons.

Theoretical and computational studies in recent years have advanced the hypothesis that movement of dislocations is a key part of the microprocess that leads up to breakdown \cite{pohjonen2011dislocation, nordlund2012defect, pohjonen2013dislocation, muszynski2015dislocation, engelberg2018stochastic}. This result supports that hypothesis. The observed superior breakdown performance of hard copper over soft copper is consistent with this hypothesis, as hard copper has a higher density of obstacles to dislocation movement, including a higher dislocation density. The observed conditioning of soft copper is also consistent with this hypothesis, as the pulsing of voltage provides a mechanism through which the dislocation density of soft copper increases, hardening it and reducing dislocation mobility. Our experiments thus provide empirical support for the aforementioned studies.

For the field of high-gradient accelerating structure development, this result suggests that thermal treatments such as those applied to the current candidates for CLIC accelerating structures \cite{wang2010fabrication} come at the cost of creating a need for a long subsequent conditioning period. The same structure performance could possibly be achieved without such a long conditioning period if structures were instead manufactured from hard copper and not subjected to treatments that soften them. In the manufacturing method currently in use, the thermal treatments are needed in order to create a contiguous structure from stacked disks with no seams, joints or other interfaces between them. However, an alternative manufacturing method has been explored in recent years: Assembly from milled semicylindrical halves \cite{catalanlasheras2016fabrication, zha2017design} (or, a variant of the same concept, quarter-cylindrical quadrants \cite{higo2008fabrication, abe2015basic}). As the halves are joined together along longitudinal planes across which no rf current flows during structure operation, the presence of an interface between the joined parts is less of an issue than for a structure assembled from stacked disks. Hence, structure assembly from milled halves requires less intensive methods for joining them. Hence, our results make the exploration of the milled halves manufacturing method even more relevant, as it might provide a way to produce a functional accelerating structure of hard copper. The procurement and testing of such a hard copper prototype structure is underway at CLIC at the time of writing.

Finally, our results could have metallurgical implications beyond the subjects of breakdown and accelerator technology. Increasing the hardness of a metal is useful for many purposes beyond breakdown, and our results suggest that such hardening could be achieved by the repeated application of surface electric fields to the metal. This might, for instance, be useful in micro-scale manufacturing where mechanical metalworking might be difficult. The utility of such a method would, however, depend on how deep below the metal surface the hardening effect reaches, and on the extent to which our results are true for other metals than copper.

\begin{acknowledgments}

We wish to extend our sincere thanks to past colleagues in the DC breakdown lab. This work was made possible by persistent long-term development of DC experimental setups and methods. Particular recognition is due Dr. Nicholas Shipman, who did a tremendous work developing the Large Electrode DC Spark System, but whose time in the lab ran out soon after having brought it to an experiment-ready state, leaving it to those of us who came after him to use it to its full effect. Special thanks are also due Dr. Vahur Zadin for his analysis of the Large Electrode DC Spark System. He identified the significance of sideways electrode misalignment, thus discovering the cause of the tendency for breakdowns to cluster around one point at the cathode edge.

\end{acknowledgments}


%
%

%


\bibliography{manuscript_bibliography}

\begin{thebibliography}{51}%
\makeatletter
\providecommand \@ifxundefined [1]{%
 \@ifx{#1\undefined}
}%
\providecommand \@ifnum [1]{%
 \ifnum #1\expandafter \@firstoftwo
 \else \expandafter \@secondoftwo
 \fi
}%
\providecommand \@ifx [1]{%
 \ifx #1\expandafter \@firstoftwo
 \else \expandafter \@secondoftwo
 \fi
}%
\providecommand \natexlab [1]{#1}%
\providecommand \enquote  [1]{``#1''}%
\providecommand \bibnamefont  [1]{#1}%
\providecommand \bibfnamefont [1]{#1}%
\providecommand \citenamefont [1]{#1}%
\providecommand \href@noop [0]{\@secondoftwo}%
\providecommand \href [0]{\begingroup \@sanitize@url \@href}%
\providecommand \@href[1]{\@@startlink{#1}\@@href}%
\providecommand \@@href[1]{\endgroup#1\@@endlink}%
\providecommand \@sanitize@url [0]{\catcode `\\12\catcode `\$12\catcode
  `\&12\catcode `\#12\catcode `\^12\catcode `\_12\catcode `\%12\relax}%
\providecommand \@@startlink[1]{}%
\providecommand \@@endlink[0]{}%
\providecommand \url  [0]{\begingroup\@sanitize@url \@url }%
\providecommand \@url [1]{\endgroup\@href {#1}{\urlprefix }}%
\providecommand \urlprefix  [0]{URL }%
\providecommand \Eprint [0]{\href }%
\providecommand \doibase [0]{http://dx.doi.org/}%
\providecommand \selectlanguage [0]{\@gobble}%
\providecommand \bibinfo  [0]{\@secondoftwo}%
\providecommand \bibfield  [0]{\@secondoftwo}%
\providecommand \translation [1]{[#1]}%
\providecommand \BibitemOpen [0]{}%
\providecommand \bibitemStop [0]{}%
\providecommand \bibitemNoStop [0]{.\EOS\space}%
\providecommand \EOS [0]{\spacefactor3000\relax}%
\providecommand \BibitemShut  [1]{\csname bibitem#1\endcsname}%
\let\auto@bib@innerbib\@empty
\bibitem [{\citenamefont {Kovermann}(2011)}]{kovermann2011comparative}%
  \BibitemOpen
  \bibfield  {author} {\bibinfo {author} {\bibfnamefont {J.}~\bibnamefont
  {Kovermann}},\ }\emph {\bibinfo {title} {Comparative studies of high-gradient
  rf and DC breakdowns}},\ \href@noop {} {Ph.D. thesis},\ \bibinfo  {school}
  {RWTH Aachen University} (\bibinfo {year} {2011})\BibitemShut {NoStop}%
\bibitem [{\citenamefont {Timk{\'o}}(2011)}]{timko2011modelling}%
  \BibitemOpen
  \bibfield  {author} {\bibinfo {author} {\bibfnamefont {H.}~\bibnamefont
  {Timk{\'o}}},\ }\emph {\bibinfo {title} {Modelling vacuum arcs: from plasma
  initiation to surface interactions}},\ \href@noop {} {Ph.D. thesis},\
  \bibinfo  {school} {University of Helsinki} (\bibinfo {year}
  {2011})\BibitemShut {NoStop}%
\bibitem [{\citenamefont {Shipman}(2015)}]{shipman2015experimental}%
  \BibitemOpen
  \bibfield  {author} {\bibinfo {author} {\bibfnamefont {N.~C.}\ \bibnamefont
  {Shipman}},\ }\emph {\bibinfo {title} {Experimental study of DC vacuum
  breakdown and application to high-gradient accelerating structures for
  CLIC}},\ \href@noop {} {Ph.D. thesis},\ \bibinfo  {school} {The University of
  Manchester, Manchester, UK} (\bibinfo {year} {2015})\BibitemShut {NoStop}%
\bibitem [{\citenamefont {Descoeudres}\ \emph
  {et~al.}(2009{\natexlab{a}})\citenamefont {Descoeudres}, \citenamefont
  {Ramsvik}, \citenamefont {Calatroni}, \citenamefont {Taborelli},\ and\
  \citenamefont {Wuensch}}]{descoeudres2009dcbreakdown}%
  \BibitemOpen
  \bibfield  {author} {\bibinfo {author} {\bibfnamefont {A.}~\bibnamefont
  {Descoeudres}}, \bibinfo {author} {\bibfnamefont {T.}~\bibnamefont
  {Ramsvik}}, \bibinfo {author} {\bibfnamefont {S.}~\bibnamefont {Calatroni}},
  \bibinfo {author} {\bibfnamefont {M.}~\bibnamefont {Taborelli}}, \ and\
  \bibinfo {author} {\bibfnamefont {W.}~\bibnamefont {Wuensch}},\ }\href@noop
  {} {\bibfield  {journal} {\bibinfo  {journal} {Physical Review Special
  Topics-Accelerators and Beams}\ }\textbf {\bibinfo {volume} {12}},\ \bibinfo
  {pages} {032001} (\bibinfo {year} {2009}{\natexlab{a}})}\BibitemShut
  {NoStop}%
\bibitem [{\citenamefont {Descoeudres}\ \emph
  {et~al.}(2009{\natexlab{b}})\citenamefont {Descoeudres}, \citenamefont
  {Djurabekova},\ and\ \citenamefont {Nordlund}}]{descoeudres2009cobalt}%
  \BibitemOpen
  \bibfield  {author} {\bibinfo {author} {\bibfnamefont {A.}~\bibnamefont
  {Descoeudres}}, \bibinfo {author} {\bibfnamefont {F.}~\bibnamefont
  {Djurabekova}}, \ and\ \bibinfo {author} {\bibfnamefont {K.}~\bibnamefont
  {Nordlund}},\ }\href@noop {} {\emph {\bibinfo {title} {DC breakdown
  experiments with cobalt electrodes}}},\ \bibinfo {type} {Tech. Rep.}\
  \bibinfo {number} {CLIC-Note-875}\ (\bibinfo  {institution} {CERN},\ \bibinfo
  {year} {2009})\BibitemShut {NoStop}%
\bibitem [{\citenamefont {Aicheler}\ \emph {et~al.}(2012)\citenamefont
  {Aicheler}, \citenamefont {Burrows}, \citenamefont {Draper}, \citenamefont
  {Garvey}, \citenamefont {Lebrun}, \citenamefont {Peach}, \citenamefont
  {Phinney}, \citenamefont {Schmickler}, \citenamefont {Schulte}, \citenamefont
  {Toge} \emph {et~al.}}]{aicheler2012cdr}%
  \BibitemOpen
  \bibfield  {author} {\bibinfo {author} {\bibfnamefont {M.}~\bibnamefont
  {Aicheler}}, \bibinfo {author} {\bibfnamefont {P.}~\bibnamefont {Burrows}},
  \bibinfo {author} {\bibfnamefont {M.}~\bibnamefont {Draper}}, \bibinfo
  {author} {\bibfnamefont {T.}~\bibnamefont {Garvey}}, \bibinfo {author}
  {\bibfnamefont {P.}~\bibnamefont {Lebrun}}, \bibinfo {author} {\bibfnamefont
  {K.}~\bibnamefont {Peach}}, \bibinfo {author} {\bibfnamefont
  {N.}~\bibnamefont {Phinney}}, \bibinfo {author} {\bibfnamefont
  {H.}~\bibnamefont {Schmickler}}, \bibinfo {author} {\bibfnamefont
  {D.}~\bibnamefont {Schulte}}, \bibinfo {author} {\bibfnamefont
  {N.}~\bibnamefont {Toge}},  \emph {et~al.},\ }\href@noop {} {\emph {\bibinfo
  {title} {A Multi-TeV linear collider based on CLIC technology: CLIC
  Conceptual Design Report}}},\ \bibinfo {type} {Tech. Rep.}\ \bibinfo {number}
  {CERN-2012-007}\ (\bibinfo  {institution} {CERN},\ \bibinfo {year}
  {2012})\BibitemShut {NoStop}%
\bibitem [{\citenamefont {Behnke}\ \emph {et~al.}(2013)\citenamefont {Behnke},
  \citenamefont {Brau}, \citenamefont {Foster}, \citenamefont {Fuster},
  \citenamefont {Harrison}, \citenamefont {McEwan~Paterson}, \citenamefont
  {Peskin}, \citenamefont {Stanitzki}, \citenamefont {Walker},\ and\
  \citenamefont {Yamamoto}}]{behnke2013ilctdr}%
  \BibitemOpen
  \bibfield  {author} {\bibinfo {author} {\bibfnamefont {T.}~\bibnamefont
  {Behnke}}, \bibinfo {author} {\bibfnamefont {J.~E.}\ \bibnamefont {Brau}},
  \bibinfo {author} {\bibfnamefont {B.}~\bibnamefont {Foster}}, \bibinfo
  {author} {\bibfnamefont {J.}~\bibnamefont {Fuster}}, \bibinfo {author}
  {\bibfnamefont {M.}~\bibnamefont {Harrison}}, \bibinfo {author}
  {\bibfnamefont {J.}~\bibnamefont {McEwan~Paterson}}, \bibinfo {author}
  {\bibfnamefont {M.}~\bibnamefont {Peskin}}, \bibinfo {author} {\bibfnamefont
  {M.}~\bibnamefont {Stanitzki}}, \bibinfo {author} {\bibfnamefont
  {N.}~\bibnamefont {Walker}}, \ and\ \bibinfo {author} {\bibfnamefont
  {H.}~\bibnamefont {Yamamoto}},\ }\href@noop {} {\emph {\bibinfo {title} {The
  International Linear Collider Technical Design Report}}},\ \bibinfo {type}
  {Tech. Rep.}\ (\bibinfo  {institution} {International Linear Collider
  Organisation},\ \bibinfo {year} {2013})\BibitemShut {NoStop}%
\bibitem [{eur(2013)}]{europeanstrategy2013update}%
  \BibitemOpen
  \href@noop {} {\enquote {\bibinfo {title} {The european strategy for particle
  physics update 2013},}\ }\bibinfo {howpublished} {CERN-Council-S/106}
  (\bibinfo {year} {2013}),\ \bibinfo {note}
  {https://cds.cern.ch/record/1567258/files/esc-e-106.pdf}\BibitemShut
  {NoStop}%
\bibitem [{\citenamefont {Krammer}(2013)}]{krammer2013update}%
  \BibitemOpen
  \bibfield  {author} {\bibinfo {author} {\bibfnamefont {M.}~\bibnamefont
  {Krammer}},\ }\href@noop {} {\bibfield  {journal} {\bibinfo  {journal}
  {Physica Scripta}\ }\textbf {\bibinfo {volume} {2013}},\ \bibinfo {pages}
  {014019} (\bibinfo {year} {2013})}\BibitemShut {NoStop}%
\bibitem [{\citenamefont {Aleksan}(2013)}]{aleksan2013future}%
  \BibitemOpen
  \bibfield  {author} {\bibinfo {author} {\bibfnamefont {R.}~\bibnamefont
  {Aleksan}},\ }\href@noop {} {\bibfield  {journal} {\bibinfo  {journal}
  {Physica Scripta}\ }\textbf {\bibinfo {volume} {2013}},\ \bibinfo {pages}
  {014016} (\bibinfo {year} {2013})}\BibitemShut {NoStop}%
\bibitem [{\citenamefont {Nakada}(2015)}]{nakada2015european}%
  \BibitemOpen
  \bibfield  {author} {\bibinfo {author} {\bibfnamefont {T.}~\bibnamefont
  {Nakada}},\ }\href@noop {} {\bibfield  {journal} {\bibinfo  {journal}
  {Nuclear Physics News}\ }\textbf {\bibinfo {volume} {25}} (\bibinfo {year}
  {2015})}\BibitemShut {NoStop}%
\bibitem [{\citenamefont {Dolgashev}\ and\ \citenamefont
  {Raubenheimer}(2004)}]{dolgashev2004simulation}%
  \BibitemOpen
  \bibfield  {author} {\bibinfo {author} {\bibfnamefont {V.~A.}\ \bibnamefont
  {Dolgashev}}\ and\ \bibinfo {author} {\bibfnamefont {T.}~\bibnamefont
  {Raubenheimer}},\ }in\ \href@noop {} {\emph {\bibinfo {booktitle}
  {Proceedings of LINAC2004, L\"ubeck, Germany}}}\ (\bibinfo {year} {2004})\
  pp.\ \bibinfo {pages} {395--398}\BibitemShut {NoStop}%
\bibitem [{\citenamefont {Palaia}(2013)}]{palaia2013beam}%
  \BibitemOpen
  \bibfield  {author} {\bibinfo {author} {\bibfnamefont {A.}~\bibnamefont
  {Palaia}},\ }\emph {\bibinfo {title} {Beam Momentum Changes due to Discharges
  in High-gradient Accelerator Structures}},\ \href@noop {} {Ph.D. thesis},\
  \bibinfo  {school} {Uppsala University} (\bibinfo {year} {2013})\BibitemShut
  {NoStop}%
\bibitem [{\citenamefont {Brown}\ \emph {et~al.}(1989)\citenamefont {Brown},
  \citenamefont {Geschonke}, \citenamefont {Henke},\ and\ \citenamefont
  {Wilson}}]{brown1989status}%
  \BibitemOpen
  \bibfield  {author} {\bibinfo {author} {\bibfnamefont {P.}~\bibnamefont
  {Brown}}, \bibinfo {author} {\bibfnamefont {G.}~\bibnamefont {Geschonke}},
  \bibinfo {author} {\bibfnamefont {H.}~\bibnamefont {Henke}}, \ and\ \bibinfo
  {author} {\bibfnamefont {I.}~\bibnamefont {Wilson}},\ }in\ \href@noop {}
  {\emph {\bibinfo {booktitle} {Particle Accelerator Conference, 1989.
  Accelerator Science and Technology., Proceedings of the 1989 IEEE}}}\
  (\bibinfo {organization} {IEEE},\ \bibinfo {year} {1989})\ pp.\ \bibinfo
  {pages} {1128--1130}\BibitemShut {NoStop}%
\bibitem [{\citenamefont {Adolphsen}\ \emph {et~al.}(2001)\citenamefont
  {Adolphsen}, \citenamefont {Baumgartner}, \citenamefont {Jobe}, \citenamefont
  {Le~Pimpec}, \citenamefont {Loewen}, \citenamefont {McCormick}, \citenamefont
  {Ross}, \citenamefont {Smith}, \citenamefont {Wang},\ and\ \citenamefont
  {Higo}}]{adolphsen2001processing}%
  \BibitemOpen
  \bibfield  {author} {\bibinfo {author} {\bibfnamefont {C.}~\bibnamefont
  {Adolphsen}}, \bibinfo {author} {\bibfnamefont {W.}~\bibnamefont
  {Baumgartner}}, \bibinfo {author} {\bibfnamefont {K.}~\bibnamefont {Jobe}},
  \bibinfo {author} {\bibfnamefont {F.}~\bibnamefont {Le~Pimpec}}, \bibinfo
  {author} {\bibfnamefont {R.}~\bibnamefont {Loewen}}, \bibinfo {author}
  {\bibfnamefont {D.}~\bibnamefont {McCormick}}, \bibinfo {author}
  {\bibfnamefont {M.}~\bibnamefont {Ross}}, \bibinfo {author} {\bibfnamefont
  {T.}~\bibnamefont {Smith}}, \bibinfo {author} {\bibfnamefont
  {J.}~\bibnamefont {Wang}}, \ and\ \bibinfo {author} {\bibfnamefont
  {T.}~\bibnamefont {Higo}},\ }in\ \href@noop {} {\emph {\bibinfo {booktitle}
  {Proceedings of the 2001 Particle Accelerator Conference}}},\ Vol.~\bibinfo
  {volume} {1}\ (\bibinfo {organization} {IEEE},\ \bibinfo {address} {Chicago,
  Illinois},\ \bibinfo {year} {2001})\ pp.\ \bibinfo {pages}
  {478--480}\BibitemShut {NoStop}%
\bibitem [{\citenamefont {Adolphsen}(2003)}]{adolphsen2003normal}%
  \BibitemOpen
  \bibfield  {author} {\bibinfo {author} {\bibfnamefont {C.}~\bibnamefont
  {Adolphsen}},\ }\href@noop {} {\emph {\bibinfo {title} {Normal-conducting RF
  structure test facilities and results}}},\ \bibinfo {type} {Tech. Rep.}\
  \bibinfo {number} {SLAC-PUB-9906}\ (\bibinfo  {institution} {SLAC},\ \bibinfo
  {year} {2003})\BibitemShut {NoStop}%
\bibitem [{\citenamefont {Rodriguez}\ \emph {et~al.}(2007)\citenamefont
  {Rodriguez}, \citenamefont {Aksakal}, \citenamefont {Arnau-Izquierdo},
  \citenamefont {Corsini}, \citenamefont {D{\"o}bert}, \citenamefont {Fandos},
  \citenamefont {Grudiev}, \citenamefont {Johnson}, \citenamefont {Mete},
  \citenamefont {Nergiz} \emph {et~al.}}]{rodriguez2007_30ghz}%
  \BibitemOpen
  \bibfield  {author} {\bibinfo {author} {\bibfnamefont {J.}~\bibnamefont
  {Rodriguez}}, \bibinfo {author} {\bibfnamefont {H.}~\bibnamefont {Aksakal}},
  \bibinfo {author} {\bibfnamefont {G.}~\bibnamefont {Arnau-Izquierdo}},
  \bibinfo {author} {\bibfnamefont {R.}~\bibnamefont {Corsini}}, \bibinfo
  {author} {\bibfnamefont {S.}~\bibnamefont {D{\"o}bert}}, \bibinfo {author}
  {\bibfnamefont {R.}~\bibnamefont {Fandos}}, \bibinfo {author} {\bibfnamefont
  {A.}~\bibnamefont {Grudiev}}, \bibinfo {author} {\bibfnamefont
  {M.}~\bibnamefont {Johnson}}, \bibinfo {author} {\bibfnamefont
  {{\"O}.}~\bibnamefont {Mete}}, \bibinfo {author} {\bibfnamefont
  {Z.}~\bibnamefont {Nergiz}},  \emph {et~al.},\ }\href@noop {} {\emph
  {\bibinfo {title} {30 GHz High-gradient accelerating structure test
  results}}},\ \bibinfo {type} {Tech. Rep.}\ \bibinfo {number} {CLIC-Note-719}\
  (\bibinfo  {institution} {CERN},\ \bibinfo {year} {2007})\BibitemShut
  {NoStop}%
\bibitem [{\citenamefont {Catalan-Lasheras}\ \emph {et~al.}(2014)\citenamefont
  {Catalan-Lasheras}, \citenamefont {Degiovanni}, \citenamefont {Dobert},
  \citenamefont {Farabolini}, \citenamefont {Kovermann}, \citenamefont
  {McMonagle}, \citenamefont {Rey}, \citenamefont {Syratchev}, \citenamefont
  {Timeo}, \citenamefont {Wuensch} \emph
  {et~al.}}]{catalanlasheras2014experience}%
  \BibitemOpen
  \bibfield  {author} {\bibinfo {author} {\bibfnamefont {N.}~\bibnamefont
  {Catalan-Lasheras}}, \bibinfo {author} {\bibfnamefont {A.}~\bibnamefont
  {Degiovanni}}, \bibinfo {author} {\bibfnamefont {S.}~\bibnamefont {Dobert}},
  \bibinfo {author} {\bibfnamefont {W.}~\bibnamefont {Farabolini}}, \bibinfo
  {author} {\bibfnamefont {J.}~\bibnamefont {Kovermann}}, \bibinfo {author}
  {\bibfnamefont {G.}~\bibnamefont {McMonagle}}, \bibinfo {author}
  {\bibfnamefont {S.}~\bibnamefont {Rey}}, \bibinfo {author} {\bibfnamefont
  {I.}~\bibnamefont {Syratchev}}, \bibinfo {author} {\bibfnamefont
  {L.}~\bibnamefont {Timeo}}, \bibinfo {author} {\bibfnamefont
  {W.}~\bibnamefont {Wuensch}},  \emph {et~al.},\ }in\ \href@noop {} {\emph
  {\bibinfo {booktitle} {Proceedings of IPAC2014}}}\ (\bibinfo {address}
  {Dresden, Germany},\ \bibinfo {year} {2014})\ pp.\ \bibinfo {pages}
  {2288--2290}\BibitemShut {NoStop}%
\bibitem [{\citenamefont {Degiovanni}\ \emph
  {et~al.}(2016{\natexlab{a}})\citenamefont {Degiovanni}, \citenamefont
  {Wuensch},\ and\ \citenamefont {Navarro}}]{degiovanni2016conditioning}%
  \BibitemOpen
  \bibfield  {author} {\bibinfo {author} {\bibfnamefont {A.}~\bibnamefont
  {Degiovanni}}, \bibinfo {author} {\bibfnamefont {W.}~\bibnamefont {Wuensch}},
  \ and\ \bibinfo {author} {\bibfnamefont {J.}~\bibnamefont {Navarro}},\
  }\href@noop {} {\bibfield  {journal} {\bibinfo  {journal} {Physical Review
  Accelerators and Beams}\ }\textbf {\bibinfo {volume} {19}},\ \bibinfo {pages}
  {032001} (\bibinfo {year} {2016}{\natexlab{a}})}\BibitemShut {NoStop}%
\bibitem [{\citenamefont {Matsumoto}\ \emph {et~al.}(2011)\citenamefont
  {Matsumoto}, \citenamefont {Abe}, \citenamefont {Higashi}, \citenamefont
  {Higo},\ and\ \citenamefont {Du}}]{matsumoto2011high}%
  \BibitemOpen
  \bibfield  {author} {\bibinfo {author} {\bibfnamefont {S.}~\bibnamefont
  {Matsumoto}}, \bibinfo {author} {\bibfnamefont {T.}~\bibnamefont {Abe}},
  \bibinfo {author} {\bibfnamefont {Y.}~\bibnamefont {Higashi}}, \bibinfo
  {author} {\bibfnamefont {T.}~\bibnamefont {Higo}}, \ and\ \bibinfo {author}
  {\bibfnamefont {Y.}~\bibnamefont {Du}},\ }\href@noop {} {\bibfield  {journal}
  {\bibinfo  {journal} {Nuclear Instruments and Methods in Physics Research
  Section A: Accelerators, Spectrometers, Detectors and Associated Equipment}\
  }\textbf {\bibinfo {volume} {657}},\ \bibinfo {pages} {160} (\bibinfo {year}
  {2011})}\BibitemShut {NoStop}%
\bibitem [{\citenamefont {Degiovanni}\ \emph {et~al.}(2014)\citenamefont
  {Degiovanni}, \citenamefont {Dobert}, \citenamefont {Farabolini},
  \citenamefont {Grudiev}, \citenamefont {Kovermann}, \citenamefont
  {Montesinos}, \citenamefont {Riddone}, \citenamefont {Syratchev},
  \citenamefont {Wegner}, \citenamefont {Wuensch} \emph
  {et~al.}}]{degiovanni2014high}%
  \BibitemOpen
  \bibfield  {author} {\bibinfo {author} {\bibfnamefont {A.}~\bibnamefont
  {Degiovanni}}, \bibinfo {author} {\bibfnamefont {S.}~\bibnamefont {Dobert}},
  \bibinfo {author} {\bibfnamefont {W.}~\bibnamefont {Farabolini}}, \bibinfo
  {author} {\bibfnamefont {A.}~\bibnamefont {Grudiev}}, \bibinfo {author}
  {\bibfnamefont {J.}~\bibnamefont {Kovermann}}, \bibinfo {author}
  {\bibfnamefont {E.}~\bibnamefont {Montesinos}}, \bibinfo {author}
  {\bibfnamefont {G.}~\bibnamefont {Riddone}}, \bibinfo {author} {\bibfnamefont
  {I.}~\bibnamefont {Syratchev}}, \bibinfo {author} {\bibfnamefont
  {R.}~\bibnamefont {Wegner}}, \bibinfo {author} {\bibfnamefont
  {W.}~\bibnamefont {Wuensch}},  \emph {et~al.},\ }in\ \href@noop {} {\emph
  {\bibinfo {booktitle} {Proceedings of IPAC2014, Dresden, Germany}}}\
  (\bibinfo {year} {2014})\ pp.\ \bibinfo {pages} {2285--2287}\BibitemShut
  {NoStop}%
\bibitem [{\citenamefont {Zennaro}\ \emph {et~al.}(2017)\citenamefont
  {Zennaro}, \citenamefont {Argyropoulos}, \citenamefont {Blumer},
  \citenamefont {Bopp}, \citenamefont {Catal{\'a}n~Lasheras}, \citenamefont
  {Esperante~Pereira}, \citenamefont {Garvey}, \citenamefont {Grudiev},
  \citenamefont {Lucas}, \citenamefont {McMonagle} \emph
  {et~al.}}]{zennaro2017high}%
  \BibitemOpen
  \bibfield  {author} {\bibinfo {author} {\bibfnamefont {R.}~\bibnamefont
  {Zennaro}}, \bibinfo {author} {\bibfnamefont {T.}~\bibnamefont
  {Argyropoulos}}, \bibinfo {author} {\bibfnamefont {H.}~\bibnamefont
  {Blumer}}, \bibinfo {author} {\bibfnamefont {M.}~\bibnamefont {Bopp}},
  \bibinfo {author} {\bibfnamefont {N.}~\bibnamefont {Catal{\'a}n~Lasheras}},
  \bibinfo {author} {\bibfnamefont {D.}~\bibnamefont {Esperante~Pereira}},
  \bibinfo {author} {\bibfnamefont {T.}~\bibnamefont {Garvey}}, \bibinfo
  {author} {\bibfnamefont {A.}~\bibnamefont {Grudiev}}, \bibinfo {author}
  {\bibfnamefont {T.}~\bibnamefont {Lucas}}, \bibinfo {author} {\bibfnamefont
  {G.}~\bibnamefont {McMonagle}},  \emph {et~al.},\ }in\ \href@noop {} {\emph
  {\bibinfo {booktitle} {8th Int. Particle Accelerator Conf.(IPAC'17),
  Copenhagen, Denmark, 14{\^a} 19 May, 2017}}}\ (\bibinfo {organization}
  {JACOW, Geneva, Switzerland},\ \bibinfo {year} {2017})\ pp.\ \bibinfo {pages}
  {4318--4320}\BibitemShut {NoStop}%
\bibitem [{\citenamefont {Wuensch}(2017)}]{wuensch2017high}%
  \BibitemOpen
  \bibfield  {author} {\bibinfo {author} {\bibfnamefont {W.}~\bibnamefont
  {Wuensch}},\ }in\ \href@noop {} {\emph {\bibinfo {booktitle} {28th Linear
  Accelerator Conf.(LINAC'16), East Lansing, MI, USA, 25-30 September 2016}}}\
  (\bibinfo {organization} {JACOW, Geneva, Switzerland},\ \bibinfo {year}
  {2017})\ pp.\ \bibinfo {pages} {368--373}\BibitemShut {NoStop}%
\bibitem [{\citenamefont {Grudiev}\ and\ \citenamefont
  {Wuensch}(2010)}]{grudiev2010design}%
  \BibitemOpen
  \bibfield  {author} {\bibinfo {author} {\bibfnamefont {A.}~\bibnamefont
  {Grudiev}}\ and\ \bibinfo {author} {\bibfnamefont {W.}~\bibnamefont
  {Wuensch}},\ }in\ \href@noop {} {\emph {\bibinfo {booktitle} {Proceedings of
  LINAC2010}}}\ (\bibinfo {address} {Tsukuba, Japan},\ \bibinfo {year} {2010})\
  pp.\ \bibinfo {pages} {211--213}\BibitemShut {NoStop}%
\bibitem [{\citenamefont {Grudiev}\ \emph {et~al.}(2009)\citenamefont
  {Grudiev}, \citenamefont {Calatroni},\ and\ \citenamefont
  {Wuensch}}]{grudiev2009new}%
  \BibitemOpen
  \bibfield  {author} {\bibinfo {author} {\bibfnamefont {A.}~\bibnamefont
  {Grudiev}}, \bibinfo {author} {\bibfnamefont {S.}~\bibnamefont {Calatroni}},
  \ and\ \bibinfo {author} {\bibfnamefont {W.}~\bibnamefont {Wuensch}},\
  }\href@noop {} {\bibfield  {journal} {\bibinfo  {journal} {Physical Review
  Special Topics-Accelerators and Beams}\ }\textbf {\bibinfo {volume} {12}},\
  \bibinfo {pages} {102001} (\bibinfo {year} {2009})}\BibitemShut {NoStop}%
\bibitem [{\citenamefont {Jacewicz}\ \emph {et~al.}(2011)\citenamefont
  {Jacewicz}, \citenamefont {Ruber}, \citenamefont {Ziemann},\ and\
  \citenamefont {Kovermann}}]{jacewicz2011instrumentation}%
  \BibitemOpen
  \bibfield  {author} {\bibinfo {author} {\bibfnamefont {M.}~\bibnamefont
  {Jacewicz}}, \bibinfo {author} {\bibfnamefont {R.}~\bibnamefont {Ruber}},
  \bibinfo {author} {\bibfnamefont {V.}~\bibnamefont {Ziemann}}, \ and\
  \bibinfo {author} {\bibfnamefont {J.}~\bibnamefont {Kovermann}}\ }(\bibinfo
  {year} {2011})\BibitemShut {NoStop}%
\bibitem [{\citenamefont {Woolley}\ \emph
  {et~al.}(2015{\natexlab{a}})\citenamefont {Woolley}, \citenamefont {Wegner},
  \citenamefont {Grudiev}, \citenamefont {Wuensch}, \citenamefont {Apsimon},
  \citenamefont {Dexter}, \citenamefont {Burt}, \citenamefont {Ambattu},\ and\
  \citenamefont {Syratchev}}]{woolley2015high}%
  \BibitemOpen
  \bibfield  {author} {\bibinfo {author} {\bibfnamefont {B.}~\bibnamefont
  {Woolley}}, \bibinfo {author} {\bibfnamefont {R.}~\bibnamefont {Wegner}},
  \bibinfo {author} {\bibfnamefont {A.}~\bibnamefont {Grudiev}}, \bibinfo
  {author} {\bibfnamefont {W.}~\bibnamefont {Wuensch}}, \bibinfo {author}
  {\bibfnamefont {R.}~\bibnamefont {Apsimon}}, \bibinfo {author} {\bibfnamefont
  {A.}~\bibnamefont {Dexter}}, \bibinfo {author} {\bibfnamefont
  {G.}~\bibnamefont {Burt}}, \bibinfo {author} {\bibfnamefont {P.~K.}\
  \bibnamefont {Ambattu}}, \ and\ \bibinfo {author} {\bibfnamefont
  {I.}~\bibnamefont {Syratchev}},\ }in\ \href@noop {} {\emph {\bibinfo
  {booktitle} {Proceedings of IPAC2015}}}\ (\bibinfo {address} {Richmond,
  Virginia},\ \bibinfo {year} {2015})\BibitemShut {NoStop}%
\bibitem [{\citenamefont {Woolley}\ \emph
  {et~al.}(2015{\natexlab{b}})\citenamefont {Woolley}, \citenamefont {Dexter},
  \citenamefont {Syratchev},\ and\ \citenamefont {Burt}}]{woolley2015high_phd}%
  \BibitemOpen
  \bibfield  {author} {\bibinfo {author} {\bibfnamefont {B.}~\bibnamefont
  {Woolley}}, \bibinfo {author} {\bibfnamefont {A.}~\bibnamefont {Dexter}},
  \bibinfo {author} {\bibfnamefont {I.}~\bibnamefont {Syratchev}}, \ and\
  \bibinfo {author} {\bibfnamefont {G.}~\bibnamefont {Burt}},\ }\emph {\bibinfo
  {title} {High power X-band RF test stand development and high power testing
  of the CLIC crab cavity}},\ \href@noop {} {Ph.D. thesis},\ \bibinfo  {school}
  {Lancaster University} (\bibinfo {year} {2015}{\natexlab{b}})\BibitemShut
  {NoStop}%
\bibitem [{\citenamefont {Kildemo}(2004)}]{kildemo2004new}%
  \BibitemOpen
  \bibfield  {author} {\bibinfo {author} {\bibfnamefont {M.}~\bibnamefont
  {Kildemo}},\ }\href@noop {} {\bibfield  {journal} {\bibinfo  {journal}
  {Nuclear Instruments and Methods in Physics Research Section A: Accelerators,
  Spectrometers, Detectors and Associated Equipment}\ }\textbf {\bibinfo
  {volume} {530}},\ \bibinfo {pages} {596} (\bibinfo {year}
  {2004})}\BibitemShut {NoStop}%
\bibitem [{\citenamefont {Rajamaki}\ \emph {et~al.}(2014)\citenamefont
  {Rajamaki}, \citenamefont {Sjobak},\ and\ \citenamefont
  {Wuensch}}]{rajamaki2014breakdown}%
  \BibitemOpen
  \bibfield  {author} {\bibinfo {author} {\bibfnamefont {R.}~\bibnamefont
  {Rajamaki}}, \bibinfo {author} {\bibfnamefont {K.~N.}\ \bibnamefont
  {Sjobak}}, \ and\ \bibinfo {author} {\bibfnamefont {W.}~\bibnamefont
  {Wuensch}},\ }\href@noop {} {\emph {\bibinfo {title} {Breakdown localization
  in the fixed gap system}}},\ \bibinfo {type} {Tech. Rep.}\ \bibinfo {number}
  {CLIC-Note-719}\ (\bibinfo  {institution} {CERN},\ \bibinfo {year}
  {2014})\BibitemShut {NoStop}%
\bibitem [{\citenamefont {Soares}\ \emph {et~al.}(2012)\citenamefont {Soares},
  \citenamefont {Wuensch}, \citenamefont {Kovermann}, \citenamefont
  {Calatroni},\ and\ \citenamefont {Barnes}}]{soares2012pulsegenerator}%
  \BibitemOpen
  \bibfield  {author} {\bibinfo {author} {\bibfnamefont {R.}~\bibnamefont
  {Soares}}, \bibinfo {author} {\bibfnamefont {W.}~\bibnamefont {Wuensch}},
  \bibinfo {author} {\bibfnamefont {J.}~\bibnamefont {Kovermann}}, \bibinfo
  {author} {\bibfnamefont {S.}~\bibnamefont {Calatroni}}, \ and\ \bibinfo
  {author} {\bibfnamefont {M.}~\bibnamefont {Barnes}},\ }in\ \href@noop {}
  {\emph {\bibinfo {booktitle} {Proceedings of IPAC2012}}},\ Vol.\ \bibinfo
  {volume} {1205201}\ (\bibinfo {year} {2012})\ p.\ \bibinfo {pages}
  {THPPC061}\BibitemShut {NoStop}%
\bibitem [{\citenamefont {Wuensch}\ \emph {et~al.}(2017)\citenamefont
  {Wuensch}, \citenamefont {Degiovanni}, \citenamefont {Calatroni},
  \citenamefont {Korsb{\"a}ck}, \citenamefont {Djurabekova}, \citenamefont
  {Rajam{\"a}ki},\ and\ \citenamefont {Giner-Navarro}}]{wuensch2017statistics}%
  \BibitemOpen
  \bibfield  {author} {\bibinfo {author} {\bibfnamefont {W.}~\bibnamefont
  {Wuensch}}, \bibinfo {author} {\bibfnamefont {A.}~\bibnamefont {Degiovanni}},
  \bibinfo {author} {\bibfnamefont {S.}~\bibnamefont {Calatroni}}, \bibinfo
  {author} {\bibfnamefont {A.}~\bibnamefont {Korsb{\"a}ck}}, \bibinfo {author}
  {\bibfnamefont {F.}~\bibnamefont {Djurabekova}}, \bibinfo {author}
  {\bibfnamefont {R.}~\bibnamefont {Rajam{\"a}ki}}, \ and\ \bibinfo {author}
  {\bibfnamefont {J.}~\bibnamefont {Giner-Navarro}},\ }\href@noop {} {\bibfield
   {journal} {\bibinfo  {journal} {Physical Review Accelerators and Beams}\
  }\textbf {\bibinfo {volume} {20}},\ \bibinfo {pages} {011007} (\bibinfo
  {year} {2017})}\BibitemShut {NoStop}%
\bibitem [{\citenamefont {Wang}\ \emph {et~al.}(2010)\citenamefont {Wang},
  \citenamefont {Lewandowski}, \citenamefont {Van~Pelt}, \citenamefont
  {Yoneda}, \citenamefont {Riddone}, \citenamefont {Gudkov}, \citenamefont
  {Higo}, \citenamefont {Takatomi} \emph {et~al.}}]{wang2010fabrication}%
  \BibitemOpen
  \bibfield  {author} {\bibinfo {author} {\bibfnamefont {J.~W.}\ \bibnamefont
  {Wang}}, \bibinfo {author} {\bibfnamefont {J.}~\bibnamefont {Lewandowski}},
  \bibinfo {author} {\bibfnamefont {J.}~\bibnamefont {Van~Pelt}}, \bibinfo
  {author} {\bibfnamefont {C.}~\bibnamefont {Yoneda}}, \bibinfo {author}
  {\bibfnamefont {G.}~\bibnamefont {Riddone}}, \bibinfo {author} {\bibfnamefont
  {D.}~\bibnamefont {Gudkov}}, \bibinfo {author} {\bibfnamefont
  {T.}~\bibnamefont {Higo}}, \bibinfo {author} {\bibfnamefont {T.}~\bibnamefont
  {Takatomi}},  \emph {et~al.},\ }in\ \href@noop {} {\emph {\bibinfo
  {booktitle} {IPAC10, THPEA064}}}\ (\bibinfo {address} {Kyoto,Japan},\
  \bibinfo {year} {2010})\BibitemShut {NoStop}%
\bibitem [{\citenamefont {Sgobba}\ and\ \citenamefont
  {Leaux}(2015)}]{cern_accelerating_structure_copper}%
  \BibitemOpen
  \bibfield  {author} {\bibinfo {author} {\bibfnamefont {S.}~\bibnamefont
  {Sgobba}}\ and\ \bibinfo {author} {\bibfnamefont {F.}~\bibnamefont {Leaux}},\
  }\href@noop {} {\emph {\bibinfo {title} {Cu-OFE bars/blanks/ingots}}},\
  \bibinfo {type} {Tech. Rep.}\ \bibinfo {number} {Technical specification No
  2001, EDMS No 790779}\ (\bibinfo  {institution} {CERN},\ \bibinfo {year}
  {2015})\ \bibinfo {note}
  {\url{https://edms.cern.ch/document/790779/6}}\BibitemShut {NoStop}%
\bibitem [{\citenamefont {Shioiri}\ \emph {et~al.}(2000)\citenamefont
  {Shioiri}, \citenamefont {Kamikawaji}, \citenamefont {Yokokura},
  \citenamefont {Kaneko}, \citenamefont {Ohshima},\ and\ \citenamefont
  {Yanabu}}]{shioiri2000dielectric}%
  \BibitemOpen
  \bibfield  {author} {\bibinfo {author} {\bibfnamefont {T.}~\bibnamefont
  {Shioiri}}, \bibinfo {author} {\bibfnamefont {T.}~\bibnamefont {Kamikawaji}},
  \bibinfo {author} {\bibfnamefont {K.}~\bibnamefont {Yokokura}}, \bibinfo
  {author} {\bibfnamefont {E.}~\bibnamefont {Kaneko}}, \bibinfo {author}
  {\bibfnamefont {I.}~\bibnamefont {Ohshima}}, \ and\ \bibinfo {author}
  {\bibfnamefont {S.}~\bibnamefont {Yanabu}},\ }in\ \href@noop {} {\emph
  {\bibinfo {booktitle} {Proceedings ISDEIV. 19th International Symposium on
  Discharges and Electrical Insulation in Vacuum}}},\ Vol.~\bibinfo {volume}
  {1}\ (\bibinfo {organization} {IEEE},\ \bibinfo {address} {Xian, China},\
  \bibinfo {year} {2000})\ pp.\ \bibinfo {pages} {17--20}\BibitemShut {NoStop}%
\bibitem [{\citenamefont {Degiovanni}\ \emph
  {et~al.}(2016{\natexlab{b}})\citenamefont {Degiovanni}, \citenamefont
  {Mouriz~Irazabal}, \citenamefont {Wegner},\ and\ \citenamefont
  {Aicheler}}]{degiovanni2016postmortem}%
  \BibitemOpen
  \bibfield  {author} {\bibinfo {author} {\bibfnamefont {A.}~\bibnamefont
  {Degiovanni}}, \bibinfo {author} {\bibfnamefont {N.}~\bibnamefont
  {Mouriz~Irazabal}}, \bibinfo {author} {\bibfnamefont {R.}~\bibnamefont
  {Wegner}}, \ and\ \bibinfo {author} {\bibfnamefont {M.}~\bibnamefont
  {Aicheler}},\ }\href@noop {} {\emph {\bibinfo {title} {Post-Mortem Analysis
  after High-Power Operation of the TD24\_R05 Tested in Xbox\_1}}},\ \bibinfo
  {type} {Tech. Rep.}\ \bibinfo {number} {CLIC-Note-1070}\ (\bibinfo
  {institution} {CERN},\ \bibinfo {year} {2016})\BibitemShut {NoStop}%
\bibitem [{\citenamefont {Nordlund}\ and\ \citenamefont
  {Djurabekova}(2012)}]{nordlund2012defect}%
  \BibitemOpen
  \bibfield  {author} {\bibinfo {author} {\bibfnamefont {K.}~\bibnamefont
  {Nordlund}}\ and\ \bibinfo {author} {\bibfnamefont {F.}~\bibnamefont
  {Djurabekova}},\ }\href@noop {} {\bibfield  {journal} {\bibinfo  {journal}
  {Physical Review Special Topics-Accelerators and Beams}\ }\textbf {\bibinfo
  {volume} {15}},\ \bibinfo {pages} {071002} (\bibinfo {year}
  {2012})}\BibitemShut {NoStop}%
\bibitem [{\citenamefont {Engelberg}\ \emph {et~al.}(2018)\citenamefont
  {Engelberg}, \citenamefont {Ashkenazy},\ and\ \citenamefont
  {Assaf}}]{engelberg2018stochastic}%
  \BibitemOpen
  \bibfield  {author} {\bibinfo {author} {\bibfnamefont {E.~Z.}\ \bibnamefont
  {Engelberg}}, \bibinfo {author} {\bibfnamefont {Y.}~\bibnamefont
  {Ashkenazy}}, \ and\ \bibinfo {author} {\bibfnamefont {M.}~\bibnamefont
  {Assaf}},\ }\href@noop {} {\bibfield  {journal} {\bibinfo  {journal}
  {Physical Review Letters}\ }\textbf {\bibinfo {volume} {120}},\ \bibinfo
  {pages} {124801} (\bibinfo {year} {2018})}\BibitemShut {NoStop}%
\bibitem [{\citenamefont {Pohjonen}\ \emph {et~al.}(2011)\citenamefont
  {Pohjonen}, \citenamefont {Djurabekova}, \citenamefont {Nordlund},
  \citenamefont {Kuronen},\ and\ \citenamefont
  {Fitzgerald}}]{pohjonen2011dislocation}%
  \BibitemOpen
  \bibfield  {author} {\bibinfo {author} {\bibfnamefont {A.}~\bibnamefont
  {Pohjonen}}, \bibinfo {author} {\bibfnamefont {F.}~\bibnamefont
  {Djurabekova}}, \bibinfo {author} {\bibfnamefont {K.}~\bibnamefont
  {Nordlund}}, \bibinfo {author} {\bibfnamefont {A.}~\bibnamefont {Kuronen}}, \
  and\ \bibinfo {author} {\bibfnamefont {S.}~\bibnamefont {Fitzgerald}},\
  }\href@noop {} {\bibfield  {journal} {\bibinfo  {journal} {Journal of Applied
  Physics}\ }\textbf {\bibinfo {volume} {110}},\ \bibinfo {pages} {023509}
  (\bibinfo {year} {2011})}\BibitemShut {NoStop}%
\bibitem [{\citenamefont {Pohjonen}\ \emph {et~al.}(2013)\citenamefont
  {Pohjonen}, \citenamefont {Parviainen}, \citenamefont {Muranaka},\ and\
  \citenamefont {Djurabekova}}]{pohjonen2013dislocation}%
  \BibitemOpen
  \bibfield  {author} {\bibinfo {author} {\bibfnamefont {A.}~\bibnamefont
  {Pohjonen}}, \bibinfo {author} {\bibfnamefont {S.}~\bibnamefont
  {Parviainen}}, \bibinfo {author} {\bibfnamefont {T.}~\bibnamefont
  {Muranaka}}, \ and\ \bibinfo {author} {\bibfnamefont {F.}~\bibnamefont
  {Djurabekova}},\ }\href@noop {} {\bibfield  {journal} {\bibinfo  {journal}
  {Journal of Applied Physics}\ }\textbf {\bibinfo {volume} {114}},\ \bibinfo
  {pages} {033519} (\bibinfo {year} {2013})}\BibitemShut {NoStop}%
\bibitem [{\citenamefont {Muszynski}(2015)}]{muszynski2015dislocation}%
  \BibitemOpen
  \bibfield  {author} {\bibinfo {author} {\bibfnamefont {J.~M.}\ \bibnamefont
  {Muszynski}},\ }\emph {\bibinfo {title} {Dislocation mechanisms and
  activation barriers of protrusion formation on a near-surface void}},\
  \href@noop {} {Master's thesis},\ \bibinfo  {school} {University of Helsinki}
  (\bibinfo {year} {2015})\BibitemShut {NoStop}%
\bibitem [{\citenamefont {Parviainen}\ \emph {et~al.}(2011)\citenamefont
  {Parviainen}, \citenamefont {Djurabekova}, \citenamefont {Timko},\ and\
  \citenamefont {Nordlund}}]{parviainen2011electronic}%
  \BibitemOpen
  \bibfield  {author} {\bibinfo {author} {\bibfnamefont {S.}~\bibnamefont
  {Parviainen}}, \bibinfo {author} {\bibfnamefont {F.}~\bibnamefont
  {Djurabekova}}, \bibinfo {author} {\bibfnamefont {H.}~\bibnamefont {Timko}},
  \ and\ \bibinfo {author} {\bibfnamefont {K.}~\bibnamefont {Nordlund}},\
  }\href@noop {} {\bibfield  {journal} {\bibinfo  {journal} {Computational
  Materials Science}\ }\textbf {\bibinfo {volume} {50}},\ \bibinfo {pages}
  {2075} (\bibinfo {year} {2011})}\BibitemShut {NoStop}%
\bibitem [{\citenamefont {Timko}\ \emph {et~al.}(2011)\citenamefont {Timko},
  \citenamefont {Matyash}, \citenamefont {Schneider}, \citenamefont
  {Djurabekova}, \citenamefont {Nordlund}, \citenamefont {Hansen},
  \citenamefont {Descoeudres}, \citenamefont {Kovermann}, \citenamefont
  {Grudiev}, \citenamefont {Wuensch} \emph {et~al.}}]{timko2011onedimensional}%
  \BibitemOpen
  \bibfield  {author} {\bibinfo {author} {\bibfnamefont {H.}~\bibnamefont
  {Timko}}, \bibinfo {author} {\bibfnamefont {K.}~\bibnamefont {Matyash}},
  \bibinfo {author} {\bibfnamefont {R.}~\bibnamefont {Schneider}}, \bibinfo
  {author} {\bibfnamefont {F.}~\bibnamefont {Djurabekova}}, \bibinfo {author}
  {\bibfnamefont {K.}~\bibnamefont {Nordlund}}, \bibinfo {author}
  {\bibfnamefont {A.}~\bibnamefont {Hansen}}, \bibinfo {author} {\bibfnamefont
  {A.}~\bibnamefont {Descoeudres}}, \bibinfo {author} {\bibfnamefont
  {J.}~\bibnamefont {Kovermann}}, \bibinfo {author} {\bibfnamefont
  {A.}~\bibnamefont {Grudiev}}, \bibinfo {author} {\bibfnamefont
  {W.}~\bibnamefont {Wuensch}},  \emph {et~al.},\ }\href@noop {} {\bibfield
  {journal} {\bibinfo  {journal} {Contributions to Plasma Physics}\ }\textbf
  {\bibinfo {volume} {51}},\ \bibinfo {pages} {5} (\bibinfo {year}
  {2011})}\BibitemShut {NoStop}%
\bibitem [{\citenamefont {Hull}\ and\ \citenamefont
  {Bacon}(2001)}]{hull2001introduction}%
  \BibitemOpen
  \bibfield  {author} {\bibinfo {author} {\bibfnamefont {D.}~\bibnamefont
  {Hull}}\ and\ \bibinfo {author} {\bibfnamefont {D.~J.}\ \bibnamefont
  {Bacon}},\ }\href@noop {} {\emph {\bibinfo {title} {Introduction to
  dislocations}}}\ (\bibinfo  {publisher} {Butterworth-Heinemann},\ \bibinfo
  {year} {2001})\BibitemShut {NoStop}%
\bibitem [{\citenamefont {Hall}(1951)}]{hall1951deformation}%
  \BibitemOpen
  \bibfield  {author} {\bibinfo {author} {\bibfnamefont {E.}~\bibnamefont
  {Hall}},\ }\href@noop {} {\bibfield  {journal} {\bibinfo  {journal}
  {Proceedings of the Physical Society. Section B}\ }\textbf {\bibinfo {volume}
  {64}},\ \bibinfo {pages} {747} (\bibinfo {year} {1951})}\BibitemShut
  {NoStop}%
\bibitem [{\citenamefont {Petch}(1953)}]{petch1953cleavage}%
  \BibitemOpen
  \bibfield  {author} {\bibinfo {author} {\bibfnamefont {N.}~\bibnamefont
  {Petch}},\ }\href@noop {} {\bibfield  {journal} {\bibinfo  {journal} {Journal
  of the Iron and Steel Institute}\ }\textbf {\bibinfo {volume} {174}},\
  \bibinfo {pages} {25} (\bibinfo {year} {1953})}\BibitemShut {NoStop}%
\bibitem [{\citenamefont {Cordero}\ \emph {et~al.}(2016)\citenamefont
  {Cordero}, \citenamefont {Knight},\ and\ \citenamefont
  {Schuh}}]{cordero2016sixdecades}%
  \BibitemOpen
  \bibfield  {author} {\bibinfo {author} {\bibfnamefont {Z.~C.}\ \bibnamefont
  {Cordero}}, \bibinfo {author} {\bibfnamefont {B.~E.}\ \bibnamefont {Knight}},
  \ and\ \bibinfo {author} {\bibfnamefont {C.~A.}\ \bibnamefont {Schuh}},\
  }\href@noop {} {\bibfield  {journal} {\bibinfo  {journal} {International
  Materials Reviews}\ }\textbf {\bibinfo {volume} {61}},\ \bibinfo {pages}
  {495} (\bibinfo {year} {2016})}\BibitemShut {NoStop}%
\bibitem [{\citenamefont {Catalan-Lasheras}\ \emph {et~al.}()\citenamefont
  {Catalan-Lasheras}, \citenamefont {Grudiev}, \citenamefont {Mcmonagle},
  \citenamefont {Syrachev}, \citenamefont {Woolley}, \citenamefont {Wuensch},
  \citenamefont {Zha}, \citenamefont {Rajamaki}, \citenamefont {Giansiracusa},
  \citenamefont {Lucas} \emph {et~al.}}]{catalanlasheras2016fabrication}%
  \BibitemOpen
  \bibfield  {author} {\bibinfo {author} {\bibfnamefont {N.}~\bibnamefont
  {Catalan-Lasheras}}, \bibinfo {author} {\bibfnamefont {A.}~\bibnamefont
  {Grudiev}}, \bibinfo {author} {\bibfnamefont {G.}~\bibnamefont {Mcmonagle}},
  \bibinfo {author} {\bibfnamefont {I.}~\bibnamefont {Syrachev}}, \bibinfo
  {author} {\bibfnamefont {B.}~\bibnamefont {Woolley}}, \bibinfo {author}
  {\bibfnamefont {W.}~\bibnamefont {Wuensch}}, \bibinfo {author} {\bibfnamefont
  {H.}~\bibnamefont {Zha}}, \bibinfo {author} {\bibfnamefont {R.}~\bibnamefont
  {Rajamaki}}, \bibinfo {author} {\bibfnamefont {P.}~\bibnamefont
  {Giansiracusa}}, \bibinfo {author} {\bibfnamefont {T.}~\bibnamefont {Lucas}},
   \emph {et~al.},\ }in\ \href@noop {} {\emph {\bibinfo {booktitle} {IPAC2016
  (not yet published)}}},\ \bibinfo {note}
  {\url{https://www.researchgate.net/publication/310747951_FABRICATION_AND_HIGH-GRADIENT_TESTING_OF_AN_ACCELERATING_STRUCTURE_MADE_FROM_MILLED_HALVES}}\BibitemShut
  {NoStop}%
\bibitem [{\citenamefont {Zha}\ and\ \citenamefont
  {Grudiev}(2017)}]{zha2017design}%
  \BibitemOpen
  \bibfield  {author} {\bibinfo {author} {\bibfnamefont {H.}~\bibnamefont
  {Zha}}\ and\ \bibinfo {author} {\bibfnamefont {A.}~\bibnamefont {Grudiev}},\
  }\href@noop {} {\bibfield  {journal} {\bibinfo  {journal} {Physical Review
  Accelerators and Beams}\ }\textbf {\bibinfo {volume} {20}},\ \bibinfo {pages}
  {042001} (\bibinfo {year} {2017})}\BibitemShut {NoStop}%
\bibitem [{\citenamefont {Higo}\ \emph {et~al.}(2008)\citenamefont {Higo},
  \citenamefont {Higashi}, \citenamefont {Kawamata}, \citenamefont {Takatomi},
  \citenamefont {Ueno}, \citenamefont {Watanabe}, \citenamefont {Yokoyama},
  \citenamefont {Grudiev}, \citenamefont {Riddone}, \citenamefont {Taborelli}
  \emph {et~al.}}]{higo2008fabrication}%
  \BibitemOpen
  \bibfield  {author} {\bibinfo {author} {\bibfnamefont {T.}~\bibnamefont
  {Higo}}, \bibinfo {author} {\bibfnamefont {Y.}~\bibnamefont {Higashi}},
  \bibinfo {author} {\bibfnamefont {H.}~\bibnamefont {Kawamata}}, \bibinfo
  {author} {\bibfnamefont {T.}~\bibnamefont {Takatomi}}, \bibinfo {author}
  {\bibfnamefont {K.}~\bibnamefont {Ueno}}, \bibinfo {author} {\bibfnamefont
  {Y.}~\bibnamefont {Watanabe}}, \bibinfo {author} {\bibfnamefont
  {K.}~\bibnamefont {Yokoyama}}, \bibinfo {author} {\bibfnamefont
  {A.}~\bibnamefont {Grudiev}}, \bibinfo {author} {\bibfnamefont
  {G.}~\bibnamefont {Riddone}}, \bibinfo {author} {\bibfnamefont
  {M.}~\bibnamefont {Taborelli}},  \emph {et~al.},\ }in\ \href@noop {} {\emph
  {\bibinfo {booktitle} {EPAC08}}}\ (\bibinfo {address} {Genoa, Italy},\
  \bibinfo {year} {2008})\BibitemShut {NoStop}%
\bibitem [{\citenamefont {Abe}\ \emph {et~al.}(2015)\citenamefont {Abe},
  \citenamefont {Arakida}, \citenamefont {Higo}, \citenamefont {Matsumoto},\
  and\ \citenamefont {Takatomi}}]{abe2015basic}%
  \BibitemOpen
  \bibfield  {author} {\bibinfo {author} {\bibfnamefont {T.}~\bibnamefont
  {Abe}}, \bibinfo {author} {\bibfnamefont {Y.}~\bibnamefont {Arakida}},
  \bibinfo {author} {\bibfnamefont {T.}~\bibnamefont {Higo}}, \bibinfo {author}
  {\bibfnamefont {S.}~\bibnamefont {Matsumoto}}, \ and\ \bibinfo {author}
  {\bibfnamefont {T.}~\bibnamefont {Takatomi}},\ }in\ \href@noop {} {\emph
  {\bibinfo {booktitle} {International Workshop on Breakdown Science and High
  Gradient Technology (HG2015}}}\ (\bibinfo {address} {Tsinghua University,
  Beijing, China},\ \bibinfo {year} {2015})\BibitemShut {NoStop}%
\end{thebibliography}%

\end{document}